\documentclass[pdflatex,sn-mathphys-num]{sn-jnl}

\usepackage{graphicx}%
\usepackage{multirow}%
\usepackage{amsmath,amssymb,amsfonts}%
\usepackage{mathrsfs}%
\usepackage[title]{appendix}%
\usepackage{xcolor}%
\usepackage{textcomp}%
\usepackage{manyfoot}%
\usepackage{booktabs}%
\usepackage[varqu]{inconsolata} 
\usepackage{hyperref}
\usepackage{xurl}
\usepackage{subcaption}
\usepackage[strict]{changepage}
\usepackage{dsfont}

\AtBeginDocument{%
}

\newcommand{\codeurl}[1]{%
  \href{#1}{{\fontfamily{zi4}\selectfont #1}}%
}

\providecommand{\argmin}{\arg\!\min}
\renewcommand{\Re}{\operatorname{Re}}
\renewcommand{\Im}{\operatorname{Im}}

\newcommand\blfootnote[1]{%
    \begingroup
    \renewcommand\thefootnote{}%
    \footnote{#1}%
    \addtocounter{footnote}{-1}%
    \endgroup
}

\usepackage[ruled,norelsize]{algorithm2e}

\usepackage{listings}%
\usepackage{array}
\usepackage{mathtools}
\usepackage{threeparttable}
\usepackage{cellspace} 
\usepackage{makecell}
\usepackage{hhline}
\usepackage{rotating}

\usepackage{pgffor}
\usepackage{alphalph} 

\theoremstyle{thmstyleone}%
\newtheorem{theorem}{Theorem}%

\theoremstyle{thmstyletwo}%

\usepackage{fancyhdr}

\fancypagestyle{figurepage}{%
  \fancyhf{} 

  \fancyfoot[L]{\hspace*{15cm}\raisebox{-6.0cm}{\thepage}}
}

\theoremstyle{thmstylethree}%

\author*[1]{\fnm{Youval} \sur{Klioui}}\email{y.klioui@tue.nl}

\affil*[1]{\orgdiv{ECE Department}, \orgname{Eindhoven University of Technology}, \orgaddress{\street{Groene Loper 3}, \city{Eindhoven}, \postcode{5612 AE}, \country{Netherlands}}}

\begin{document}
\urlstyle{tt}
\title[Article Title]{SR-TL1: A Square-Root TL1-Norm Framework for Robust SMV DoA Estimation under Highly-Coherent Dictionaries}

\abstract{This paper proposes a Square-Root Transformed $L_1$-norm ($SR\text{-}TL_{1}$) sparse recovery framework for single-measurement-vector (SMV) direction of arrival (DoA) estimation under highly-coherent overcomplete dictionaries with angular-dependent array imperfections. The proposed framework combines the square-root Least Absolute Shrinkage and Selection Operator (square-root LASSO) framework which is known to be robust against noise variance with the Transformed $L_1$-norm ($TL_1$-norm), a non-convex penalty that shows a stronger recovery performance than the classical convex $L_1$-norm under highly-coherent dictionaries. We use the Difference of Convex Algorithm (DCA) along with the Alternating Direction Method of Multipliers (ADMM) algorithm to obtain simple, closed-form update rules and provide an efficient implementation that leverages the low-rank nature of the Gram matrix of the dictionary so as to obtain a computational complexity of at most $\mathcal{O}(MN)$ per iteration where $M$ is the array size and $N$ is the length of the dictionary. We additionally provide a convergence guarantee for the DCA iterates of $SR\text{-}TL_{1}$. Experimental verification of the proposed framework shows a lower sensitivity of the regularization hyperparameter to the noise variance level and a competitive recovery performance compared to state-of-the-art baselines.\blfootnote{The repository for replicating the results reported here can be found at: \codeurl{https://github.com/youvalklioui/sr-tl1-doa}}}

\keywords{Compressed Sensing, DoA estimation, Square-root LASSO, $TL_{1}$-Norm }

\maketitle

\section{Introduction}
\label{introduction}
Direction of arrival (DoA) estimation using only a single measurement vector (SMV) and under angular-dependent array imperfections remains a challenging problem in the array signal processing community. The presence of angular-dependent errors renders state-of-the-art methods that are based on a gridless approach such as Atomic Norm Minimization (ANM) \cite{anm}\cite{ram} ineffective since the semidefinite problem (SDP) that underlies these methods fundamentally assumes an ideal array manifold with no imperfections whatsoever. Recent work introduced by Chen et al. mitigates this issue partly by considering an improved version of ANM termed GP-ANM \cite{anm-error}, in which the authors assumed unknown mismatch errors and formulated a corresponding SDP that incorporates the model mismatch. Despite the improved performance compared to classical ANM, their method only considers the simple case of angular-independent array imperfections and without mutual coupling effects, which is typically not the case in practice. Additionally, the proposed GP-ANM works under the assumption of multiple measurement vectors (MMV), an assumption which, naturally, does not hold under the single-snapshot setting. Older classical methods such as Multiple Signal Classification (MUSIC) \cite{smusic} or Estimation of Signal Parameters via Rotational Invariance Techniques (ESPRIT) \cite{esprit} show severe performance degradation in the single snapshot case with array imperfections primarily because the spatial smoothing pre-processing method \cite{SS}, which is necessary under the SMV case, is ineffective since the assumption of sub-array shift invariance does not hold as each element of the array has its own antenna pattern. Furthermore, spatial smoothing is generally not possible for sparse arrays since in this case too the sub-array invariance property is not satisfied, an undesirable limitation since in practice sparse arrays help in reducing the effect of mutual coupling between antennas as well as increasing the array aperture, and hence the achievable resolution. Maximum Likelihood Estimation (MLE)\cite{mle} is also typically inapplicable under the array imperfections setting since the objective function in MLE is, similarly to ANM, typically formulated with the assumption of an ideal array manifold free of any imperfections. MLE also typically requires the prior knowledge of the number of sources present in the measurement vector, which is hard to obtain with only a single snapshot. Grid-based methods such as Sparse Bayesian Learning (SBL)\cite{sbl1}\cite{sbl2} and the $L_{1}$-norm based Least Absolute Shrinkage and Selection Operator (LASSO) framework \cite{lasso}, on the other hand, naturally accommodate the angular-dependent array imperfections since a given column of the dictionary used in the sparse recovery problem formulation corresponds to the value of the real array manifold recorded at a given angle. Although SBL performs well in the high signal-to-noise ratio (SNR) regime, it comes at a non-negligible computational cost since each iteration of the algorithm requires an inversion of a size $M$ matrix, where $M$ is the array size. On the other hand, $L_{1}$-norm sparse recovery, despite not being as computationally intensive as SBL, suffers from poor performance when the coherence of the dictionary increases, which is typically the case when doing high resolution DoA estimation since the angular step size used to sample the array manifold is small. Moreover, a small angular step size for sampling the array manifold is also desirable since this will typically minimize the grid mismatch effect \cite{mismatch}. Another shortcoming of the LASSO is the requirement to manually tune the $L_{1}$-norm regularization hyperparameter, which typically depends on the noise variance, a quantity that is hard to estimate under a single snapshot setting. The aforementioned issue was addressed by Belloni et al. through the so-called square-root LASSO framework \cite{sqlasso} wherein it was shown that using an $L_{2}$-norm in the data fitting term of the LASSO, as opposed to using the classical \emph{squared} $L_{2}$-norm, allowed for a sparse recovery framework where the hyperparameter of the $L_{1}$-norm becomes almost independent from the value of the noise variance. One additional benefit of the square-root LASSO is the convex nature of the optimization problem since the $L_{2}$-norm is a convex penalty, allowing for the existence of a global minimum.

\par Recently, there has been increased research in the field of non-convex sparse recovery as these methods showed stronger recovery performance under fewer measurements as well as increased robustness against highly-coherent dictionaries than the classical $L_{1}$-norm. Among the non-convex penalties, the transformed $L_{1}$-norm ($TL_1$-norm) \cite{tl1_norm} was shown to offer one of the most robust recovery performance for a wide range of highly-coherent dictionaries including highly correlated Gaussian dictionaries as well as oversampled Discrete Fourier Transform (DFT) dictionaries. The $TL_1$-norm, an elementwise sparsity penalty, is governed by a hyperparameter $\alpha$ whereby as $\alpha$ approaches zero, the behavior of the $TL_1$-norm approaches that of the $L_0$-norm in promoting sparser solutions and decreasing amplitude bias. On the other hand, the $TL_{1}$-norm, similarly to the $L_{1}$-norm in the LASSO framework, nevertheless requires the adjustment of a regularizer $\kappa$ whose optimal value depends on the noise variance level, which is again hard to estimate \emph{a priori} from a single measurement vector and even harder under low SNR settings. This work addresses this shortcoming and the main contributions are summarized as follows:
\begin{itemize}
\item We introduce a non-convex sparse recovery framework termed Square-Root $TL_1$ ($SR\text{-}TL_{1}$) that shows a lower sensitivity of the regularization hyperparameter of the non-convex penalty with respect to the noise variance.

\item We use the DCA and ADMM algorithms to provide simple closed-form update rules for the non-convex optimization problem. We additionally provide a convergence guarantee of the DCA sequence of iterates to a stationary point.
\item We provide an efficient implementation of the DCA subproblem ADMM updates that leverages the low rank nature of the Gram matrix of the dictionary so as to achieve a computational complexity of at most $\mathcal{O}(MN)$ per iteration, where $M$ is the array size and $N$ the dictionary length.
\item The recovery performance of the proposed framework is benchmarked against other baselines such as SBL, square-root LASSO, and other state-of-the-art sparse recovery methods and is shown to exhibit a competitive performance.
\end{itemize}

The rest of the paper is organized as follows. Section \ref{signalmodel} sets up the signal model where we introduce angular-dependent array imperfections. Section \ref{lasso-sqlasso} reviews the classical $L_{1}$-norm based LASSO along with the square-root LASSO framework. Section \ref{tl1_recovery} provides a brief review of the $TL_1$ sparse recovery method. Section \ref{sr_tl1_recovery} covers the proposed framework and provides an efficient implementation thereof along with a convergence guarantee. Section \ref{performance_evaluation} provides an extensive performance characterization of the proposed framework against other state-of-the-art methods with respect to the detection rate $P_{d}$, the angular root-mean squared error (RMSE), and the false alarm rate $P_{fa}$ across varying noise variance levels and under a highly-coherent dictionary setting and angular-dependent array imperfections. Section \ref{conclusion} provides a conclusion.\\

\par \emph{\underline{Notation:}} $\mathbb{C}$, $\mathbb{R}^{+}$, $\mathbb{N}$ denote the set of complex, real positive and natural numbers, respectively. Matrices are denoted by bold uppercase letters ($\mathbf{A}$), and vectors by bold lowercase letters ($\mathbf{a}$).  $|z|$ and $\textrm{arg}(z)$ denote the magnitude and argument, respectively, of the complex number $z$. $(.)^{H}$ denotes the Hermitian transpose and $\overline{(.)}$ denotes the complex conjugate. $\textrm{diag}(\mathbf{v})$ denotes the diagonal matrix obtained from $\mathbf{v}$. $\Re(\mathbf{x})$ and $\Im(\mathbf{x})$ denote the real and imaginary part, respectively, of $\mathbf{x}\in \mathbb{C}^{N}$. The subdifferential of a function $f$ is denoted by $\partial f$. $ \mathcal{CN}(\mathbf{0},\sigma^{2}\mathbf{I})$ denotes the multivariate complex normal distribution with mean $\mathbf{0}$ and covariance matrix $\sigma^{2}\mathbf{I}$. $\mathbf{x}\odot\mathbf{y}$ and $\langle \mathbf{x}, \mathbf{y} \rangle = \mathbf{x}^{H}\mathbf{y}$ denote the element-wise and inner product, respectively, of $\mathbf{x}$ and $\mathbf{y}$. $|\mathbf{x}|$ denotes the elementwise amplitude of $\mathbf{x}$, and $\mathbf{x}/\mathbf{y}$ denotes the elementwise division of $\mathbf{x}$ by $\mathbf{y}$. $\mathbf{I}_{N}$ denotes the size $N$ identity matrix, and $\mathbf{1}$ and $\mathbf{0}$ denote the ones and null vector, respectively. $\max(\mathbf{x}, \mathbf{y})$ with $\mathbf{x},\mathbf{y} \in \mathbb{R}^{N}$ is the elementwise max operator. 

\section{Signal Model}
\label{signalmodel}
\begin{figure}
	\centering
	\includegraphics[width=8cm, height = 3.2cm]{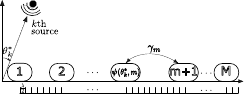}
	\caption{Array with $M$ elements located on a $\lambda/2$ grid. The gain-phase imperfection complex coefficient $\boldsymbol{\psi}(\theta_{k}^{*}, m)$ is assumed to be angular-dependent whereas the mutual coupling $\gamma_{m}$ between element $m$ and $m+1$ is not.}
	\label{array}
\end{figure}

We consider an $M$-element sparse array whose elements are positioned on a $\lambda/2$ grid, i.e. the coordinate of the $m$-th element is given by $p_{m} = k_{m}\lambda/2$ with $k_{m} \in \mathbb{N}$ as shown in Fig. \ref{array}. The signal model under the SMV setting for $K$ sources is given by
\begin{align}
\mathbf{y} = \sum_{k=1}^{K} r_{k} \mathbf{a}(\theta_{k}^{*}) + \mathbf{n},
\label{signal_model}
\end{align}
where $r_{k} \in \mathbb{C}$ is the amplitude of the source located at angular position $\theta_{k}^{*}$, $\mathbf{n} \sim 
\mathcal{CN}(\mathbf{0},\sigma^{2}\mathbf{I}_{M})$ is additive white noise, and $\mathbf{a}(\theta_{k}^{*})$ is the array steering vector at angular position $\theta_{k}^{*}$ which we model as 
\begin{align}
\mathbf{a}(\theta_{k}^{*}) = \mathbf{\Gamma} (\boldsymbol{\psi}(\theta_{k}^{*}) \odot \mathbf{a}_{i}(\theta_{k}^{*})),
\label{real_manifold}
\end{align}
where $\mathbf{a}_{i}(\theta_{k}^{*}) \in \mathbb{C}^{M}$ is the ideal array steering vector given by 
\begin{align}
\mathbf{a}_{i}(\theta_{k}^{*}, m) = \exp(-j2\pi k_{m}\sin(\theta_{k}^{*})/2),
\label{array_manifold}
\end{align}
and $\boldsymbol{\psi}(\theta_{k}^{*})$ represents the angular-dependent gain-phase imperfections vector
at angle $\theta_{k}^{*}$ where $|\boldsymbol{\psi}(\theta_{k}^{*}, m)|$ and $\mathrm{arg}(\boldsymbol{\psi}(\theta_{k}^{*}, m))$ quantify the amount of gain and  phase mismatch, respectively, ascribed to the $m$-th array element, and $\mathbf{\Gamma} \in \mathbb{C}^{M\times M}$ represents the mutual coupling matrix which we assume has the following first-order model form
\begin{align}
\mathbf{\Gamma}=
\begin{bmatrix}
1 & \gamma_{1}& 0& 0 & \hdots & 0 \\
\gamma_{1} & 1 & \gamma_{2}& 0 & \hdots & 0 \\
0 & \gamma_{2}& 1& \gamma_{3} & \hdots& 0 \\
\vdots & \ddots& \ddots& \ddots & \ddots &\vdots \\
0 & \hdots& 0 & \gamma_{M-2} & 1 &\gamma_{M-1} \\
0 & \hdots & 0 & 0 & \gamma_{M-1} & 1 \\
\end{bmatrix},
\label{mutual_coupling_matrix}
\end{align}
where $\gamma_{m}$ is the mutual coupling coefficient between element $m$ and $m+1$, and the strength of the mutual couplings is given by $|\gamma_{m}|$.
\section{LASSO and Square-Root LASSO}
\label{lasso-sqlasso}
\subsection{LASSO}
In this section we first briefly review the $L_{1}$-norm based LASSO framework that will serve as an introduction as well as baseline reference for the $TL_1$-norm sparse recovery framework that will be presented later. In DoA estimation under the LASSO framework, we define an $N$-point uniform angular grid $\mathcal{G}_{\theta} = \{\theta_{1}, \theta_{2}, \hdots, \theta_{N}\}$ at which the real array manifold $\mathbf{a}(\theta)$ is sampled where $\theta_{n} = \theta_{1} +  (n-1)\Delta\theta$ and $\Delta\theta$ is the angular step size, leading to a dictionary $\mathbf{A} \in \mathbb{C}^{M\times N}$ given by 
\begin{align}
\mathbf{A} = [\mathbf{a}(\theta_{1}), \mathbf{a}(\theta_{2}), \hdots, \mathbf{a}(\theta_{N})].
\label{dictionary}
\end{align}
Given the measurement vector $\mathbf{y}$, the goal is the recovery of its sparse representation with respect to the dictionary $\mathbf{A}$. More specifically, we assume that the true angles $\{\theta_{k}^{*}\}_{k=1}^{K}$ defined in \eqref{signal_model} belong to the angular grid $\mathcal{G}_{\theta}$ we defined, i.e. $\{\theta_{k}^{*}\}_{k=1}^{K} \subset \mathcal{G}_{\theta}$, and we assume the following representation to be true 
\begin{align}
\mathbf{y} = \mathbf{A}\mathbf{r} + \mathbf{n}, 
\label{forward_model_0}
\end{align}
where $\mathbf{r} \in \mathbb{C}^{N\times1}$ is a sparse vector that contains the amplitudes $\{r_{k}\}_{k=1}^{K}$ and whose support indicates the angular positions $\{\theta_{k}^{*}\}_{k=1}^{K}$. The sparsity assumption here specifically translates to the condition $K<<N$. Furthermore, the forward model can be written under the following equivalent form where the dictionary has unit norm columns,
\begin{align}
\mathbf{y} = \mathbf{A} \mathbf{R}^{-1} \mathbf{R}\mathbf{r} + \mathbf{n} = \mathbf{D} \mathbf{c} + \mathbf{n},
\label{forward_model}
\end{align}
where
\begin{align}
\begin{cases}
\mathbf{R}= \textrm{diag}(||\mathbf{a}(\theta_{1})||_{2}, ||\mathbf{a}(\theta_{2})||_{2}, \hdots, ||\mathbf{a}(\theta_{N})||_{2} ) \\
\mathbf{D} = \mathbf{A}\mathbf{R}^{-1}  \\
\mathbf{c} = \mathbf{R}\mathbf{r}.
\end{cases}
\label{unit_norm_dictionary}
\end{align}
We can therefore recover an estimate $\hat{\mathbf{r}}$ of the amplitude vector $\mathbf{r}$ from an estimate $\hat{\mathbf{c}}$ of $\mathbf{c}$  directly through $\hat{\mathbf{r}} = \mathbf{R}^{-1} \hat{\mathbf{c}}$. In what follows, $\mathbf{d}(\theta_{n})$ will denote the $n$-th column of $\mathbf{D}$. Under the $L_{1}$ sparse recovery framework, an estimate $\hat{\mathbf{c}}
$ of $\mathbf{c}$ is obtained by solving the following optimization problem
\begin{align}
\hat{\mathbf{c}} = \argmin_{\mathbf{x}} {\frac{1}{2}||\mathbf{y}-\mathbf{D}\mathbf{x}||_{2}^{2} + \kappa||\mathbf{x}||_{1}},
\label{lasso}
\end{align}
where $||\mathbf{x}||_{1} = \sum_{n=1}^{N}|\mathbf{x}(n)|$ is the $L_{1}$-norm that enforces the sparsity criterion and $\kappa$ is a regularization hyperparameter that controls the level of sparsity in the solution and whose value depends on the noise variance level $\sigma ^{2}$. Since the objective in $\eqref{lasso}$ is convex in addition to the $L_{1}$-norm having a well-known closed-form proximal operator \cite{moreau}, we can, under global convergence guarantees, directly apply the proximal gradient descent method \cite{boyd}, which is also known as the Iterative Shrinkage and Thresholding Algorithm (ISTA)\cite{ista} as well as its Nesterov-accelerated variant \cite{nesterov}, commonly known as the Fast Iterative Shrinkage and Thresholding Algorithm (FISTA)\cite{fista}. The simplicity of the two aforementioned algorithms made them highly popular in the compressed sensing community. ISTA, for instance, amounts to repeating the following simple mapping given an initial vector $\mathbf{x}^{(0)}$
\begin{align}
	\mathbf{x}^{(l+1)} = S((\mathbf{I}_{N} - \mu \mathbf{D}^{H}\mathbf{D})\,\mathbf{x}^{(l)} + \mu \mathbf{D}^{H}\mathbf{y};\mu\kappa), \label{ista}
\end{align}
where 
\begin{align}
S(\mathbf{x};\beta) \triangleq \frac{\mathbf{x}}{|\mathbf{x}|} \odot \max(|\mathbf{x}|-\beta\mathbf{1},\mathbf{0})
\label{soft_thresh}
\end{align} 
denotes the proximal operator for the $L_{1}$-norm, which is also known as the soft-thresholding operator, and where $\mu = 1/\nu_{max}(\mathbf{D})^{2}$ is the step size \cite{ista} with $\nu_{max}(\mathbf{D})$ being the largest singular value of $\mathbf{D}$.  

\par Despite its simplicity, the $L_{1}$ framework suffers from two major shortcomings. The first relates to the issue of sparse recovery using a highly-coherent dictionary.  As was stated before, the model in \eqref{forward_model} assumes the true angles lie exactly on the angular grid, but this assumption is rarely satisfied in practice since the true angles can in effect assume any value from a continuous range, this leads to an issue known as the grid mismatch effect \cite{mismatch} and typically leads to a degradation in the sparse recovery performance. In order to mitigate this, the density of the angular grid can be increased by decreasing the angular step size $\Delta\theta$, but this inevitably also leads to an increase in the mutual coherence $\mu(\mathbf{D})$ of the dictionary $\mathbf{D}$ which effectively quantifies how similar the columns that makeup the dictionary $\mathbf{D}$ are and is explicitly given by 
\begin{align}
\mu (\mathbf{D}) \triangleq \max_{n1,n2} \frac{|\mathbf{d}(\theta_{n_{1}})^{H}\mathbf{d}(\theta_{n_{2}})|}{||\mathbf{d}(\theta_{n_{1}})||_{2}||\mathbf{d}(\theta_{n_{2}})||_{2}}, 
\end{align}
and where values of $\mu (\mathbf{D})$ close to 1 indicate a highly-coherent dictionary which, similarly to the grid mismatch effect, decreases the sparse recovery performance \cite{coherence1}\cite{coherence2}. The second shortcoming for the classical LASSO framework is the strong dependence of the hyperparameter $\kappa$ on the noise variance, which is challenging to estimate from a single measurement vector setting.
\subsection{Square-Root LASSO}
As mentioned earlier, the value of the regularization parameter $\kappa$ for the $L_{1}$-norm used in the LASSO objective function in \eqref{lasso} depends on the noise variance $\sigma ^{2}$, whereby a higher noise variance would require a higher $\kappa$ so as to suppress more spurious noise peaks in the reconstructed spectrum. A global optimizer $\hat{\mathbf{c}}$ of the LASSO objective function in \eqref{lasso} satisfies the optimality condition given by \cite{boyd}
\begin{align}
\mathbf{0} \in  \left. \partial \left( \frac{1}{2} \Vert{}\mathbf{y} - \mathbf{D}\mathbf{x}\Vert{}_2^2 + \kappa \Vert{}\mathbf{x}\Vert{}_1 \right) \right\vert{}_{\mathbf{x} = \hat{\mathbf{c}}},
\end{align}
which can be expressed component-wise as 
\begin{align}
|\mathbf{d}_{n}^{H}(\mathbf{y}-\mathbf{D}\hat{\mathbf{c}})|  &\le  \kappa, \quad n = 1,\hdots, N \label{kappa_inequality}.
\end{align}
If we now consider the simple case where the observation is pure noise, i.e. $\mathbf{y} = \mathbf{n}$, and hence $\mathbf{c} = \mathbf{0}$, a sensible choice for $\kappa$ would force a null output with high probability for a given $\sigma^2$. We therefore have that $\hat{\mathbf{c}}=\mathbf{0}$ is the unique global optimizer if the following strict inequalities hold with high probability
\begin{align}
|\mathbf{d}_{n}^{H}\mathbf{n}|  <  \kappa, \quad n = 1,\hdots, N \label{kappa_inequality2},
\end{align}
which implies
\begin{align}
||\mathbf{D}^{H}\mathbf{n}||_{\infty} < \kappa,
\end{align}
where $||\mathbf{D}^{H}\mathbf{n}||_{\infty} = \max (|\mathbf{d}_{1}^{H}\mathbf{n}|, \hdots, |\mathbf{d}_{N}^{H}\mathbf{n}|)$. If we denote by $\pmb{\epsilon} \sim \mathcal{CN}(\mathbf{0}, \mathbf{I})$ so that $\mathbf{n} = \sigma \pmb{\epsilon}$, then the aforementioned inequality can be stated as 
\begin{align}
\sigma||\mathbf{D}^{H}\pmb{\epsilon}||_{\infty} < \kappa,
\label{threshold_condition}
\end{align}
where the quantity $||\mathbf{D}^{H}\pmb{\epsilon}||_{\infty}$ is independent of $\sigma$. Thus the minimum value of the regularization parameter $\kappa_{min} = t\sigma||\mathbf{D}^{H}\pmb{\epsilon}||_{\infty}$, with $t>1$, scales linearly with the noise standard deviation.  As mentioned earlier, determining this threshold is difficult in practice when only a single measurement vector is available for estimating the noise variance. To circumvent this issue, Belloni et al. proposed the square-root LASSO framework \cite{sqlasso} wherein we solve the following optimization problem
\begin{align}
\hat{\mathbf{c}} = \argmin_{\mathbf{x}} {||\mathbf{y}-\mathbf{D}\mathbf{x}||_{2} + \kappa||\mathbf{x}||_{1}}.
\label{sqlasso}
\end{align}
It can be shown that using the $L_{2}$-norm rather than the squared $L_{2}$-norm for the residual term removes the dependence of $\kappa$ on the noise variance.  We can note again that the objective function in \eqref{sqlasso} is still convex, and thus the same optimality condition for the global minimum applies, namely
\begin{align}
\mathbf{0} \in  \left. \partial \big(||\mathbf{y}-\mathbf{D}\mathbf{x}||_{2} +\kappa||\mathbf{x}||_{1} \big) \right\vert{}_{\mathbf{x} = \hat{\mathbf{c}}},
\label{optimality_sqlasso}
\end{align}
which can be expressed component-wise as 
\begin{align}
\frac{|\mathbf{d}_{n}^{H}(\mathbf{y}-\mathbf{D}\hat{\mathbf{c}})|}{||\mathbf{y}-\mathbf{D}\hat{\mathbf{c}}||_{2}}  &\le  \kappa, \quad n = 1,\hdots, N.
\end{align}
Under the same previous scenario, for $\mathbf{y}=\mathbf{n}$, $\hat{\mathbf{c}}=\mathbf{0}$ will be the global optimizer if the following inequality holds with high probability
\begin{align}
\frac{|\mathbf{d}_{n}^{H}\mathbf{n}|}{||\mathbf{n}||_{2}}  &<  \kappa, \quad n = 1,\hdots, N,
\end{align}
from which 
\begin{align}
\left\|\frac{\mathbf{D}^{H}\mathbf{n}}{||\mathbf{n}||_{2}} \right\|_{\infty} < \kappa \implies \left\|\frac{\mathbf{D}^{H}\pmb{\epsilon}}{||\pmb{\epsilon}||_{2}} \right\|_{\infty} < \kappa, 
\label{threshold_sqlasso}
\end{align}
where we note that the left hand side in \eqref{threshold_sqlasso} is independent of $\sigma$. Moreover, we also have 
\begin{align}
\frac{||\mathbf{D}^{H}\pmb{\epsilon}||_{\infty}}{\sqrt{M}} \lesssim \kappa,
\end{align} 
under the approximation $||\pmb{\epsilon}||_{2} \approx \sqrt{M}$. Thus, under a purely noisy observation, the minimum threshold under square-root LASSO is given by approximately $\kappa_{min} \approx t ||\mathbf{D}^{H}\pmb{\epsilon}||_{\infty} / \sqrt{M}$ and is independent of the noise variance level. We refer the reader to \cite{sqlasso} for a treatment of the more general case wherein the authors demonstrate that the minimum $\kappa_{min}$ is still independent of $\sigma^2$. 
\section{Classical $TL_1$-Norm Recovery}
\label{tl1_recovery}
\begin{figure}[t] 
	\centering
	\includegraphics[height=5cm,width=7cm]{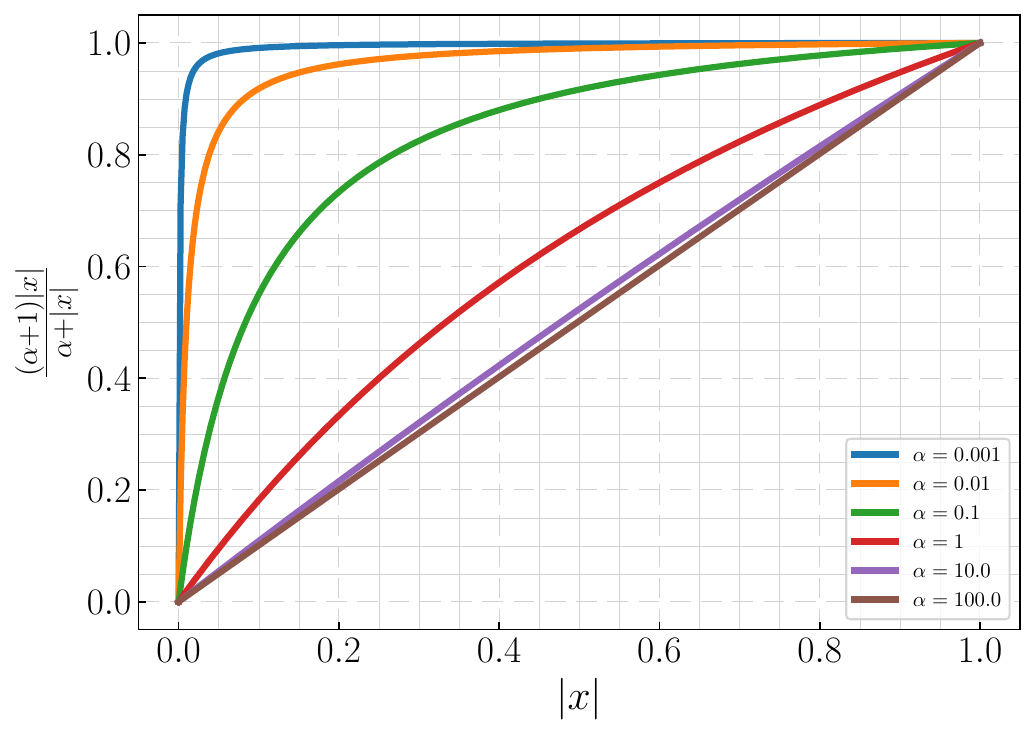}
	\caption{Plot of $r(x;\alpha) = (\alpha + 1)|x|/(\alpha + |x|)$ for $|x| \in [0,1]$ for different values of $\alpha$. For smaller values of $\alpha$ we see a saturation behavior that approaches that of the $L_{0}$-norm whereby the amplitude of a component contributes a fixed amount to the total penalty $R(\mathbf{x};\alpha) = \sum_{n=1}^{N}r(\mathbf{x}(n);\alpha)$ once the amplitude exceeds a certain threshold.}
    \label{penalty_tl1}
\end{figure}

\begin{figure}[t] 
	\centering
	\includegraphics[height=5cm,width=7cm]{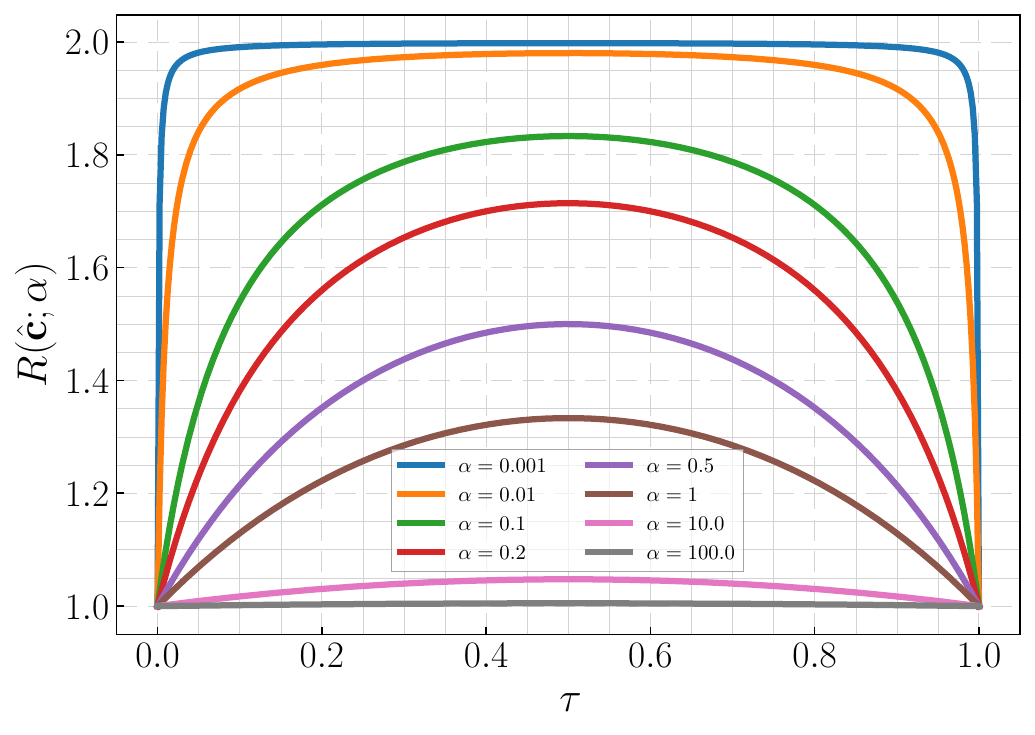}
	\caption{Plot of the $TL_1$-norm $R(\hat{\mathbf{c}};\alpha)$ for a two-component sparse vector $\hat{\mathbf{c}}(n) =  \tau c_{0}\delta(n-n_{0})+(1-\tau)c_{0}\delta(n-(n_{0}+1))$ as a function of $\tau$ for different values of $\alpha$ with $|c_{0}|=1$. For both very small and very large values of $\alpha$, the $TL_1$-norm becomes insensitive to the way with which the amplitude is split between the two atoms.}
    \label{penalty_tl1_coherence}
\end{figure}

The $TL_{1}$-norm, introduced by Zhang et al. \cite{tl1_norm}, is a non-convex sparsity penalty that was shown to exhibit a robust recovery performance under highly-coherent dictionaries of various types compared to the $L_1$ norm. Formally, the classical $TL_{1}$ sparse recovery framework is given by

\begin{align}
\hat{\mathbf{c}} = \argmin_{\mathbf{x}} {\frac{1}{2}||\mathbf{y}-\mathbf{D}\mathbf{x}||_{2}^{2} + \kappa R(\mathbf{x}; \alpha)},
\label{tl1}
\end{align}
where $R(\mathbf{x}; \alpha)$ is the $TL_1$ regularizer given by \cite{tl1_norm}\cite{tl1_regularizer}
\begin{align}
R(\mathbf{x}; \alpha) = \sum_{n=1}^{N} \frac{(\alpha + 1)|x(n)|}{\alpha + |x(n)|},
\label{tl1_regularizer}
\end{align}
and where $\alpha>0$ is a hyperparameter. It can be shown that the penalty function in \eqref{tl1_regularizer} exhibits both unbiasedness and sparsity properties \cite{tl1_regularizer}. More specifically, we can see as shown in Fig. \ref{penalty_tl1} that the mapping given by
\begin{align}
r(x;\alpha) = \frac{(\alpha + 1)|x|}{\alpha + |x|},
\label{unit_penalty_tl1}
\end{align}
for small values of $\alpha$ exhibits a saturation behavior when a component's amplitude becomes large enough. In other words, once the magnitude of a given component $\mathbf{x}(n)$ exceeds a certain threshold, which depends on the value of $\alpha$, its contribution to the total penalty $R_{\alpha}(\mathbf{x})$ remains constant. This behavior mimics that of the $L_{0}$ norm, which only adds up the number of non-zero components in a vector regardless of their amplitude. In fact, since we have
\begin{align}
\lim_{\alpha \rightarrow 0} r(x;\alpha) = \lim_{\alpha \rightarrow 0} \frac{(\alpha + 1)|x|}{\alpha + |x|} = \begin{cases}
  1 & \text{if } x \neq 0 \\
  0 & \text{if } x = 0,
\end{cases}
\end{align}
i.e. $\lim\limits_{\alpha \rightarrow 0} r(x;\alpha) = \mathds{1}(x)$, it follows that
\begin{align}
\lim_{\alpha \rightarrow 0} R_{\alpha}(\mathbf{x}) = \sum_{n=1}^{N} \lim_{\alpha \rightarrow 0} r_{\alpha}(\mathbf{x}(n)) = \sum_{n=1}^{N} \mathds{1}(\mathbf{x}(n)),
\end{align}
which corresponds to the $L_{0}$ norm of $\mathbf{x}$. On the other hand, we also have
\begin{align}
\lim_{\alpha \rightarrow +\infty} r(x;\alpha) = \lim_{\alpha \rightarrow +\infty} \frac{(1 + 1/ \alpha)|x|}{1 + |x|/\alpha} = |x|,
\end{align}
from which 
\begin{align}
\lim_{\alpha \rightarrow +\infty} R(\mathbf{x};\alpha) = \sum_{n=1}^{N} \lim_{\alpha \rightarrow +\infty} r(\mathbf{x}(n);\alpha) = \sum_{n=1}^{N} |\mathbf{x}(n)|,
\end{align}
which corresponds to the $L_{1}$-norm of $\mathbf{x}$. Thus the $TL_1$-norm effectively provides an interpolation between the $L_1$ and the $L_0$-norm as $\alpha$ varies in $(0, +\infty)$. Small values of $\alpha$ are therefore desirable in that regard since the behavior of the $TL_1$-norm in that regime approaches that of the $L_0$-norm. On the other hand, this also has the negative effect of a reduction in the penalizing effect this norm has over mutually coherent atoms. If we consider a single-component solution given by $\mathbf{c}(n) = c_{0}\delta(n-n_{0})$ along with a two-component estimate of the form  $\hat{\mathbf{c}}(n) = \tau c_{0}\delta(n-n_{0})+(1-\tau)c_{0}\delta(n-(n_{0}+1))$ with $0<\tau<1$, then the quadratic regularizer in both \eqref{lasso} and \eqref{tl1}, ignoring the constant term, reduces to
\begin{align}
Q(\hat{\mathbf{c}}; \tau) = \frac{1}{2}||\mathbf{D}\hat{\mathbf{c}}||_{2}^{2}  -\Re \langle \mathbf{y}, \mathbf{D}\hat{\mathbf{c}} \rangle, 
\end{align}
where $\mathbf{D}\hat{\mathbf{c}} = \tau\mathbf{d}(\theta_{n_{0}})+(1-\tau)\mathbf{d}(\theta_{n_{0}+1})$. Under a highly-coherent dictionary setting formed using a small angular step size $\Delta\theta$, we will have the approximation $ \mathbf{d}(\theta_{n_{0}+1}) \approx \mathbf{d}(\theta_{n_{0}})$ and thus 
\begin{align}
Q(\hat{\mathbf{c}}; \tau) \approx  \frac{1}{2}||\mathbf{d}(\theta_{n_{0}})||_{2}^{2}  -\Re \langle \mathbf{y}, \mathbf{d}(\theta_{n_{0}}) \rangle, 
\end{align}
for any $\tau$. Thus the sparsity penalty added to $Q(\hat{\mathbf{c}}; \tau)$ must be able to penalize the splitting of an amplitude between two atoms to compensate for the lack of discriminative power over highly-coherent atoms that the quadratic regularizer exhibits. For the $L_{1}$-norm this is not the case since $||\hat{\mathbf{c}}||_{1}$ = $|\tau c_{0}|+|(1-\tau)c_{0}| = |c_{0}|$ for any $0<\tau<1$. On the other hand, the $TL_1$ regularizer is given by
\begin{align}
 R(\hat{\mathbf{c}}; \alpha) = \frac{(\alpha + 1)\tau}{\frac{\alpha}{|c_{0}|} + \tau} + \frac{(\alpha + 1)(1-\tau)}{\frac{\alpha}{|c_{0}|} + 1-\tau}.
\end{align}
As $\alpha \rightarrow 0$, we have $R(\hat{\mathbf{c}}; \alpha) \rightarrow 2$ for any $0<\tau<1$, in other words the penalty becomes insensitive to the way in which the amplitude $c_{0}$ is split between the two atoms as shown in Fig. \ref{penalty_tl1_coherence}. We can note that the same behavior also emerges when the hyperparameter $\alpha$ is very large since $R(\hat{\mathbf{c}}; \alpha) \rightarrow |c_{0}|$ for $\alpha \rightarrow +\infty$ for any $0<\tau<1$. This behavior is of course expected given that in these two limit cases, the $TL_1$-norm behaves either as the $L_{0}$-norm or the $L_{1}$-norm. Thus choosing a value of $\alpha$ that is neither too small nor too high is crucial for maintaining a robust performance in the coherent setting. For example, the authors in \cite{tl1_norm} recommend a value of $\alpha = 1$. As we can note from Fig. \ref{penalty_tl1_coherence}, for $\alpha=1$ the penalty reaches its minimum for $\tau=0$ and $\tau=1$, i.e. whenever the amplitude $|c_{0}|$ is fully ascribed to a single atom. Additionally, for $\tau=1/2$ which corresponds to the amplitude being equally split between the two coherent atoms, the $TL_{1}$ penalty reaches its maximum value. We again refer the reader to \cite{tl1_norm} for a more comprehensive treatment of the $TL_{1}$-norm.

\section{Square-Root $TL_1$-Norm Recovery}
\label{sr_tl1_recovery}
\subsection{DCA Framework and ADMM Solution}
Despite its increased robustness to coherent dictionaries and strong sparse recovery performance, the classical $TL_{1}$ framework, similarly to LASSO, still requires the manual tuning of a regularization parameter $\kappa$ that exhibits a dependency on the noise variance. Thus, motivated by the noise robustness of the square-root LASSO framework, we propose a similar optimization framework whereby we use the $L_{2}$-norm for the residual in conjunction with the $TL_{1}$ penalty, which we will refer to in what follows as $SR\text{-}TL_{1}$. The general optimization problem is given by 
\begin{align}
\mathbf{c}^{*} = \argmin_{\mathbf{x}}{}||\mathbf{y}-\mathbf{D}\mathbf{x}||_{2} + \kappa R(\mathbf{x} ;\alpha)+ \frac{\tilde{\rho}}{2}||\mathbf{x}||_{2}^{2},
\label{sqncvx}
\end{align}
where $R(\mathbf{x};\alpha)$ is defined in \eqref{tl1_regularizer} and $\tilde{\rho}>0$. The $\tilde{\rho}/2||\mathbf{x}||_{2}^{2}$ regularizer is added to guarantee the boundedness of the sequence of DCA iterates and has no material effect on the convergence of the DCA algorithm using ADMM as will be seen.  The optimization problem in \eqref{sqncvx} is non-convex in nature and is typically solved through the difference of convex algorithm (DCA) \cite{dca}. Under DCA, a non-convex objective function $h(\mathbf{x})$ is reformulated as a difference of two convex functions, known as a DC decomposition, as follows
\begin{align}
\mathbf{c}^{*} = \argmin_{\mathbf{x}}h(\mathbf{x}) = \argmin_{\mathbf{x}}f(\mathbf{x}) -g(\mathbf{x}),
\label{dca_diff}
\end{align}
where $f$ and $g$ are both convex functions. A first order linear approximation $g_{L}(\mathbf{x}; \hat{\mathbf{c}}^{(t)})$ of $g$ around the current estimate of $\mathbf{c}^{*}$ at step $t$, denoted by $\hat{\mathbf{c}}^{(t)}$, is first computed as follows
\begin{align}
g_{L}(\mathbf{x}; \hat{\mathbf{c}}^{(t)}) = g(\hat{\mathbf{c}}^{(t)}) + \Re{\langle \mathbf{v}(\hat{\mathbf{c}}^{(t)}),(\mathbf{x}-\hat{\mathbf{c}}^{(t)})}\rangle,
\label{g_l1l2}
\end{align}
where $\mathbf{v}(\hat{\mathbf{c}}^{(t)}) \in \partial g(\hat{\mathbf{c}}^{(t)}) $ denotes a subgradient \cite{boyd} of $g$ at $\hat{\mathbf{c}}^{(t)}$. The linear approximation in \eqref{g_l1l2} is then substituted into \eqref{dca_diff} and we solve the following convex optimization problem to obtain the next estimate $\hat{\mathbf{c}}^{(t+1)}$,
\begin{align}
\hat{\mathbf{c}}^{(t+1)} &= \argmin\limits_{\mathbf{x}}{f(\mathbf{x})} - g_{L}(\mathbf{x}; \hat{\mathbf{c}}^{(t)}). 
\label{l1l2_linearized}
\end{align}
The value of $\hat{\mathbf{c}}^{(t+1)}$ is then used to compute a new linear approximation $g_{L}(\mathbf{x}; \hat{\mathbf{c}}^{(t+1)})$ of $g$, and $\eqref{l1l2_linearized}$ is solved again using the updated linear approximation
$g_{L}(\mathbf{x}; \hat{\mathbf{c}}^{(t+1)})$ and so forth.

\par The function $||\mathbf{y}-\mathbf{D}\mathbf{x}||_{2}$ is convex since it is a composition of a convex function with a linear map \cite{boyd}. A DC decomposition for the objective function in \eqref{sqncvx} is therefore given by 
\begin{align}
 \begin{cases}
   f(\mathbf{x}) = ||\mathbf{y}-\mathbf{D}\mathbf{x}||_{2} + \kappa' ||\mathbf{x}||_{1} + \dfrac{\tilde{\rho}}{2}||\mathbf{x}||_{2}^{2} \\[1.5ex]
  g(\mathbf{x}) = \kappa' \sum_{n=1}^{N}  \dfrac{|\mathbf{x}(n)|^{2}}{(\alpha+|\mathbf{x}(n)|)},
\end{cases}
\label{decompositon_r2}
\end{align} 
where $\kappa' = \kappa (\alpha+1)/{\alpha}$ and we made use of the following DC decomposition of the function in \eqref{unit_penalty_tl1} \cite{tl1_norm}
\begin{align}
r(x;\alpha) = \frac{(\alpha+1)|x|}{\alpha} - \frac{(\alpha+1)|x|^{2}}{\alpha(\alpha+|x|)}.
\end{align}
The function $g: \mathbb{C}^{N}\rightarrow \mathbb{R}^{+}$ is continuously differentiable, therefore its subdifferential contains a single element and is given by $\partial g (\mathbf{x})   = \{2 \nabla g (\mathbf{x})\}$ where $\nabla g (\mathbf{x})$ denotes the gradient of $g$. Next, using Wirtinger calculus \cite{wirtinger}, we have
\begin{align}
 \nabla g (\mathbf{x}) = \frac{\partial g(\mathbf{x})}{\partial \overline{\mathbf{x}}} &= \frac{\partial }{\partial \overline{\mathbf{x}}} \bigg ( \kappa' \sum_{n=1}^{N}  \frac{\overline{\mathbf{x}(n)}\mathbf{x}(n)}{(\alpha+\sqrt{\overline{\mathbf{x}(n)}\mathbf{x}(n)})}\bigg) \nonumber\\
 &= \big(\kappa' \frac{(\alpha\mathbf{1}+|\mathbf{x}|/2)}{(\alpha\mathbf{1}+|\mathbf{x}|)^{2}}\big) \odot \mathbf{x}.
 \label{subgradient_tl1norm}
\end{align}
Ignoring the constant terms, the resulting DCA subproblem at step $t$ for \eqref{sqncvx} is thus given by 
\begin{align}
\hat{\mathbf{c}}^{(t+1)} = \argmin\limits_{\mathbf{x}}{}f(\mathbf{x})- 2\Re{\langle\nabla g(\hat{\mathbf{c}}^{(t)}),
 \mathbf{x}}\rangle.
\label{dca_subproblem}
\end{align}
Algorithm \ref{alg1} provides a summary of the DCA for solving $SR\text{-}TL_{1}$.
\begin{algorithm}[htbp]
    \caption{Difference of Convex Algorithm (DCA) for solving $SR\text{-}TL_{1}$}
    \label{alg1}
    \KwIn{$\hat{\mathbf{c}}^{(0)}\in\mathbb{C}^{N\times1}$, $err$, $t_{max}$.}
    \While{$||\hat{\mathbf{c}}^{(t+1)}-\hat{\mathbf{c}}^{(t)}||_{2}>err$ \textrm{and} $t < t_{max}$}{

        $\hat{\mathbf{c}}^{(t+1)} = \operatorname*{argmin}\limits_{\mathbf{x}} ||\mathbf{y}-\mathbf{D}\mathbf{x}||_{2} +  \kappa' ||\mathbf{x}||_{1}+ \dfrac{\tilde{\rho}}{2}||\mathbf{x}||_{2}^{2} - 2\Re{\langle\nabla g(\hat{\mathbf{c}}^{(t)}),\mathbf{x}}\rangle$\;
    }
\end{algorithm}

The following theorem provides a convergence guarantee for DCA applied to $SR\text{-}TL_{1}$ with the proof available in Appendix \ref{proof_theorem}.
\begin{theorem}
\label{th:conv}
The DCA sequence $\{\hat{\mathbf{c}}^{(t)}\}$ solving \eqref{dca_subproblem} is bounded and globally converges to a critical point $\hat{\mathbf{c}}$ of the objective $h(\mathbf{x})=  ||\mathbf{y}-\mathbf{D}\mathbf{x}||_{2} +\frac{\tilde{\rho}}{2}||\mathbf{x}||_{2}^{2} + \kappa R(\mathbf{x};\alpha)$, and we additionally have $\mathbf{0} \in \partial h (\hat{\mathbf{c}}) $.
\end{theorem}

\par The optimization problem in \eqref{dca_subproblem} is convex and can be solved using the ADMM algorithm through the following reformulation
\begin{align}
\hat{\mathbf{c}}^{(t+1)} =\argmin\limits_{\mathbf{x}, \mathbf{z}_{1},\mathbf{z}_{2}} {}||\mathbf{y}-\mathbf{z}_{1}||_{2} + \kappa'||\mathbf{z}_{2}||_{1} + \frac{\tilde{\rho}}{2}||\mathbf{x}||_{2}^{2}   
- 2\Re{\langle\nabla g(\hat{\mathbf{c}}^{(t)}), \mathbf{x}}\rangle,
\label{dca_subproblem_admm}
\end{align}
with the constraints 
\begin{align}
\begin{cases}
\mathbf{z}_{1} = \mathbf{D}\mathbf{x} \\
\mathbf{z}_{2} = \mathbf{x}.
\end{cases}
\end{align}
The augmented Lagrangian corresponding to \eqref{dca_subproblem_admm} is given by
\begin{align}
\mathcal{L}(\mathbf{x},\mathbf{z}_{1},\mathbf{z}_{2},\mathbf{u}_{1},\mathbf{u}_{2})&=   ||\mathbf{y}-\mathbf{z}_{1}||_{2}  + \kappa'||\mathbf{z}_{2}||_{1} + \frac{\tilde{\rho}}{2}||\mathbf{x}||_{2}^{2} - 2\Re{\langle\nabla g(\hat{\mathbf{c}}^{(t)}), \mathbf{x}}\rangle  \nonumber \\ 
& +\frac{\rho_{2}}{2}||\mathbf{x}-\mathbf{z}_{2}+\mathbf{u}_{2}||_{2}^{2}  +\frac{\rho_{1}}{2}||\mathbf{D}\mathbf{x}-\mathbf{z}_{1}+\mathbf{u}_{1}||_{2}^{2},
\end{align}
with $\rho_{1},\rho_{2}>0$. The update rules are then given by
\begin{align}
&\mathbf{x}^{(l+1)}= \argmin\limits_{\mathbf{x}}{\mathcal{L}(\mathbf{x},\mathbf{z}_{1}^{(l)},\mathbf{z}_{2}^{(l)},\mathbf{u}_{1}^{(l)},\mathbf{u}_{2}^{(l)})} \label{x_update_admm} \\
&\mathbf{z}_{1}^{(l+1)}= \argmin\limits_{\mathbf{z}_{1}}{\mathcal{L}(\mathbf{x}^{(l+1)},\mathbf{z}_{1},\mathbf{z}_{2}^{(l)},\mathbf{u}_{1}^{(l)},\mathbf{u}_{2}^{(l)})} \label{z1_update_admm}\\
&\mathbf{z}_{2}^{(l+1)}= \argmin\limits_{\mathbf{z}_{2}}{\mathcal{L}(\mathbf{x}^{(l+1)},\mathbf{z}_{1}^{(l+1)},\mathbf{z}_{2},
\mathbf{u}_{1}^{(l)},\mathbf{u}_{2}^{(l)})} \label{z2_update_admm}\\
&\mathbf{u}_{1}^{(l+1)} = \mathbf{u}_{1}^{(l)} + \mathbf{D}\mathbf{x}^{(l+1)}-\mathbf{z}_{1}^{(l+1)} \\
&\mathbf{u}_{2}^{(l+1)} = \mathbf{u}_{2}^{(l)} + \mathbf{x}^{(l+1)}-\mathbf{z}_{2}^{(l+1)}.
\end{align}
The subproblems in \eqref{x_update_admm}, \eqref{z1_update_admm}, and \eqref{z2_update_admm} are convex in their respective update variable and therefore reduce to solving for their respective optimality condition. Thus \eqref{x_update_admm} amounts to solving
\begin{align}
    \frac{\partial \mathcal{L}(\mathbf{x}, \overline{\mathbf{x}})}{\partial \overline{\mathbf{x}}} = \mathbf{0},
\end{align}
from which 
\begin{align}
\mathbf{x}^{(l+1)} = \big(\rho_{1}\mathbf{D}^{H}\mathbf{D}+(\rho_{2}+\tilde{\rho})\mathbf{I}_{N}\big)^{-1}\big(\rho_{1}\mathbf{D}^{H}(\mathbf{z}_{1}^{(l)}-\mathbf{u}_{1}^{(l)}) + \rho_{2}(\mathbf{z}_{2}^{(l)}-\mathbf{u}_{2}^{(l)})+2\nabla g(\hat{\mathbf{c}}^{(t)})\big).
\label{x_update_admm_2}
\end{align}
As can be seen, adding the $\tilde{\rho}/{2}||\mathbf{x}||_{2}^{2}$ term in the objective in \eqref{sqncvx} amounts to shifting $\rho_{2}$ by $\tilde{\rho}$ in the computation of the inverse $(\rho_{1}\mathbf{D}^{H}\mathbf{D}+(\rho_{2}+\tilde{\rho})\mathbf{I}_{N})^{-1}$, thus if we set $\tilde{\rho} = \rho_{2}/10^{8}$, for example, its effect will be marginal on the $\mathbf{x}$ update while also guaranteeing the coercivity of the objective in \eqref{sqncvx}. In what follows we set $\tilde{\rho_{2}} = \rho_{2}+\tilde{\rho}$.  Similarly, defining $\mathbf{q}^{(l)} = \mathbf{D}\mathbf{x}^{(l+1)} + \mathbf{u}_1^{(l)} - \mathbf{y}$, and by using the formal definition of the proximal operator of the $L_{2}$-norm, \eqref{z1_update_admm} reduces to 
\begin{align}
\mathbf{z}_1^{(l+1)} &=  \argmin\limits_{\mathbf{z}_{1}} \dfrac{\rho_{1}}{2}||\mathbf{D}\mathbf{x}+\mathbf{u}_{1}-\mathbf{z}_{1}||_{2}^{2} + ||\mathbf{y}-\mathbf{z}_{1}||_{2}=\mathbf{y} + \max\left(0,\; 1 - \frac{1}{\rho_1 \| \mathbf{q}^{(l)} \|_2} \right) \mathbf{q}^{(l)}.
\end{align}
Finally, the solution to \eqref{z2_update_admm} is given by
\begin{align}
\mathbf{z}_2^{(l+1)} = S(\mathbf{x}^{(l+1)}+\mathbf{u}_{2}^{(l)}; \kappa'/\rho_{2})
\label{z2_update_admm_2}
\end{align}
where $S(\cdot;\beta)$, defined in \eqref{soft_thresh}, is the proximal operator of the $L_{1}$-norm.
\subsection{Efficient Implementation}

In this section we provide an efficient implementation of the matrix-vector product in \eqref{x_update_admm_2}, which is the largest bottleneck for a given ADMM update step. We begin with the observation that since $\mathbf{D}$ in \eqref{unit_norm_dictionary} is an overcomplete dictionary of rank $M_{r}\le M <<N$, we have that $\textrm{rank}(\mathbf{D}^{H}\mathbf{D})=\textrm{rank}(\mathbf{D}^{H}) = M_r$. Moreover, a compact singular value decomposition (SVD) of $\mathbf{D}$ is given by
\begin{align}
\mathbf{D}= \mathbf{U}\mathbf{\Sigma}\mathbf{V}^{H},
\end{align}
where $\mathbf{U} \in \mathbb{C}^{M \times M_{r}}$ and $\mathbf{V} \in \mathbb{C}^{N \times M_{r}}$  are the left and right, respectively, singular vectors of $\mathbf{D}$ and $\mathbf{\Sigma}=\textrm{diag}(\nu_{1},\nu_{2},\hdots,\nu_{M_{r}})$ are the singular values, where $\nu_{M_{r}}>0$ is the smallest singular value. Denoting the orthogonal complement of $\mathbf{V}$ by $\mathbf{V}_{\perp}$, we have for $\rho_{1},\tilde{\rho_2}> 0$
\begin{align}
&\rho_{1}\mathbf{D}^{H}\mathbf{D}+\tilde{\rho_{2}} \mathbf{I}_{N} = \mathbf{V}(\rho_{1}\mathbf{\Sigma}^{2})\mathbf{V}^{H} + \tilde{\rho_{2}} (\mathbf{V}\mathbf{V}^{H}+\mathbf{V}_{\perp}\mathbf{V}_{\perp}^{H}) \nonumber \\
&= \mathbf{V}(\rho_{1}\mathbf{\Sigma}^{2}+\tilde{\rho_{2}} \mathbf{I}_{M_{r}})\mathbf{V}^{H} + \tilde{\rho_{2}}\mathbf{V}_{\perp}\mathbf{V}_{\perp}^{H} \nonumber\\
&= [\mathbf{V}, \mathbf{V}_{\perp} ] \begin{bmatrix}
&\rho_{1}\mathbf{\Sigma}^{2}+\tilde{\rho_{2}} \mathbf{I}_{M_{r}}, &\mathbf{0}\\
&\mathbf{0},&\tilde{\rho_{2}} \mathbf{I}_{N-M_{r}}
\end{bmatrix}[\mathbf{V}, \mathbf{V}_{\perp} ]^{H}.
\end{align}
Thus $[\mathbf{V}, \mathbf{V}_{\perp} ]$ provides a block diagonalization for $\rho_{1}\mathbf{D}^{H}\mathbf{D}+\tilde{\rho_{2}} \mathbf{I}_{N}$. The inverse is therefore given by
\begin{align}
&(\rho_{1}\mathbf{D}^{H}\mathbf{D}+\tilde{\rho_{2}} \mathbf{I}_{N})^{-1} \nonumber\\
&=[\mathbf{V}, \mathbf{V}_{\perp} ] \begin{bmatrix}
&(\rho_{1}\mathbf{\Sigma}^{2}+\tilde{\rho_{2}} \mathbf{I}_{M_{r}})^{-1}, &\mathbf{0}\\
&\mathbf{0},&\dfrac{1}{\tilde{\rho_{2}}} \mathbf{I}_{N-M_{r}}
\end{bmatrix}[\mathbf{V}, \mathbf{V}_{\perp} ]^{H} \nonumber\\
&= \mathbf{V}(\rho_{1}\mathbf{\Sigma}^{2}+\tilde{\rho_{2}}\mathbf{I}_{{M}_{r}})^{-1}\mathbf{V}^{H}+\frac{1}{\tilde{\rho_{2}}}\mathbf{V}_{\perp}\mathbf{V}_{\perp}^{H} \nonumber\\
&= \mathbf{V}(\rho_{1}\mathbf{\Sigma}^{2}+\tilde{\rho_{2}}\mathbf{I}_{{M}_{r}})^{-1}\mathbf{V}^{H}+\frac{1}{\tilde{\rho_{2}}}(\mathbf{I}_{N}-\mathbf{V}\mathbf{V}^{H}) \nonumber \\
&= \mathbf{V}\big((\rho_{1}\mathbf{\Sigma}^{2}+\tilde{\rho_{2}}\mathbf{I}_{{M}_{r}})^{-1}-\frac{1}{\tilde{\rho_{2}}}\mathbf{I}_{M_{r}}\big)\mathbf{V}^{H}+\frac{1}{\tilde{\rho_{2}}}\mathbf{I}_{N} \nonumber\\
&=-\mathbf{\tilde{V}}\mathbf{\tilde{V}}^{H}+\frac{1}{\tilde{\rho_{2}}}\mathbf{I}_{N},
\end{align}
where $\mathbf{\tilde{V}} = \mathbf{V}\mathbf{\Lambda}^{1/2}$ and $\mathbf{\Lambda}$ is a size $M_{r}$ diagonal matrix with non-negative entries given by
\begin{align}
\mathbf{\Lambda}(m_{r}) = \frac{1}{\tilde{\rho_{2}}}-\frac{1}{\rho_{1}\nu_{m_{r}}^{2}+\tilde{\rho_{2}}},
\label{lambda_compute}
\end{align}
for $m_{r}= 1, \hdots, M_{r}$. Therefore, given a precomputed compact SVD of $\mathbf{D}$, a matrix-vector product of the form $(\rho_{1}\mathbf{D}^{H}\mathbf{D}+\tilde{\rho_{2}} \mathbf{I}_{N})^{-1}\mathbf{s}$ for $\mathbf{s}\in \mathbb{C}^{N}$ can be computed in $\mathcal{O}(NM_{r})$ operations. We provide a summary for the ADMM solution of the DCA subproblem in Algorithm \ref{alg2}. 

\begin{algorithm}[h!]
    \caption{ADMM for Solving $SR\text{-}TL_{1}$ DCA Subproblem}
    \DontPrintSemicolon 
    Input: $\mathbf{y},\mathbf{z}_{1}^{(0)},\mathbf{u}_{1}^{(0)}\in\mathbb{C}^{M\times1}$, $\hat{\mathbf{c}}^{(t)},\mathbf{z}_{2}^{(0)},\mathbf{u}_{2}^{(0)}\in\mathbb{C}^{N\times1}$, $\mathbf{D}\in\mathbb{C}^{M\times N}$, $\kappa, \rho_{1},\rho_{2}>0$, $\alpha>0$, $l_{max} \in \mathbb{N}$\;
    $\mathbf{U},\mathbf{\Sigma}, \mathbf{V}=\textrm{SVD}(\mathbf{D})$\;
	Precompute $\kappa' = \kappa (\alpha+1)/\alpha,\quad \tilde{\rho_{2}} = \rho_{2}(1+10^{-8})$ \;
    Precompute $\mathbf{\tilde{V}} = \mathbf{V}\mathbf{\Lambda}^{1/2}$ where $\mathbf{\Lambda}$ is defined in \eqref{lambda_compute}\;
    
	Evaluate $\nabla g(\hat{\mathbf{c}}^{(t)} )$ using \eqref{subgradient_tl1norm} \;
        
	\For{$l := 0$ to $l_{max}-1$}
	{
		$\begin{aligned}[t]
			\mathbf{s}^{(l)} \leftarrow & \rho_{1}\mathbf{D}^{H}(\mathbf{z}_{1}^{(l)}-\mathbf{u}_{1}^{(l)})+\rho_{2}(\mathbf{z}_{2}^{(l)}-\mathbf{u}_{2}^{(l)})  + 2\nabla g(\hat{\mathbf{c}}^{(t)})
		\end{aligned}$\;
		$\mathbf{x}^{(l+1)} \leftarrow -\mathbf{\tilde{V}}\mathbf{\tilde{V}}^{H}\mathbf{s}^{(l)}+\mathbf{s}^{(l)}/\tilde{\rho_{2}}$\;
		$\mathbf{q}^{(l)} \leftarrow \mathbf{u}_1^{(l)} + \mathbf{D}\mathbf{x}^{(l+1)} - \mathbf{y}$\;
		$\mathbf{z}_1^{(l+1)} \leftarrow \mathbf{y} + \max\left(0,\; 1 - \dfrac{1}{\rho_1 \| \mathbf{q}^{(l)} \|_2} \right) \mathbf{q}^{(l)}$\;
		$\mathbf{z}_2^{(l+1)} \leftarrow S(\mathbf{x}^{(l+1)}+\mathbf{u}_{2}^{(l)}; \kappa'/\rho_{2})$\;
		$\mathbf{u}_{1}^{(l+1)} \leftarrow \mathbf{u}_{1}^{(l)} + \mathbf{D}\mathbf{x}^{(l+1)}-\mathbf{z}_{1}^{(l+1)}$\;
		$\mathbf{u}_{2}^{(l+1)} \leftarrow \mathbf{u}_{2}^{(l)} + \mathbf{x}^{(l+1)}-\mathbf{z}_{2}^{(l+1)}$\;
	}

    Output: $\hat{\mathbf{c}}^{(t+1)}= \mathbf{x}^{(l_{max})}$\;
    \label{alg2}
\end{algorithm}

\section{Performance Evaluation}
\label{performance_evaluation}
\subsection{Metrics Definitions}
\label{metrics_definition}
So as to characterize the sparse recovery performance of the proposed framework, we use the detection rate $P_{d}$, the false alarm rate $P_{fa}$, and the angular root-mean-squared error (RMSE).
\par The detection rate is computed using a two-stage detection criterion. Specifically, after recovering $\hat{\mathbf{r}}$ from $\hat{\mathbf{c}}$ as described in \eqref{forward_model}, we perform a peak extraction using the magnitude spectrum $|\hat{\mathbf{r}}|$ so as to obtain a new vector $\hat{\mathbf{r}}_{pk}$.  For a given ground truth source $r_{k}$  located at an angular position $\theta_{k}^{*}$,  we first identify the set of angles 
\begin{align}
	\mathcal{S}^{1}_{\theta_{k}^{*}}=\{\theta_{k_{1}}, \theta_{k_{2}}, \ldots, \theta_{k_{E(k)}}\} \subseteq \mathcal{G}_{\theta},
\end{align}
from the support of $\hat{\mathbf{r}}_{pk}$ that satisfy a proximity criterion with respect to the true angle $\theta_{k}^{*}$,
\begin{align}
	|\theta_{k_{e}}-\theta_{k}^{*}| \le \delta\theta, \qquad e = 1, 2, \ldots, E(k).
    \label{detect_1}
\end{align}
Second, we retain only the subset of angles $\mathcal{S}^{2}_{\theta_{k}^{*}} \subseteq \mathcal{S}^{1}_{\theta_{k}^{*}}$  at which the ratio between the amplitude of the estimated spectrum and the ground-truth amplitude exceeds a predefined threshold $0<\zeta_{1} \le 1$, i.e. for each $\theta_{k_{f}} \in \mathcal{S}^{2}_{\theta_{k}^{*}}$ we have
\begin{align}
	\dfrac{|\mathbf{\hat{r}}_{pk}({k_{f}})|}{|r_{k}|} \ge \zeta_{1}, \qquad f = 1, 2, \ldots, F(k).
    \label{detect_2}
\end{align}
If the set $\mathcal{S}^{2}_{\theta_{k}^{*}}$ is non-empty, then the recovery of $\theta_{k}^{*}$ is considered successful. This process is repeated for all $K$ sources present in the measurement vector $\mathbf{y}$. Defining the set indicator function as
\begin{align}
\mathds{1}(\mathcal{S})= \begin{cases}
1 \quad \textrm{if} \quad \mathcal{S} \neq \emptyset \\
0  \quad \textrm{if} \quad \mathcal{S} = \emptyset,
\end{cases}
\end{align}
along with the number of detections $T_{D} = \sum_{k=1}^{K}\mathds{1}(\mathcal{S}^{2}_{\theta_{k}^{*}})$, the  detection rate $P_{d}$ is then given by
\begin{align}
	P_{d} =\dfrac{T_{D}}{K},
\label{pd_def}
\end{align}
and the false alarm rate $P_{fa}$ for a given threshold $\zeta_{2}>0$ is defined as 
\begin{align}
	P_{fa} =\dfrac{|| \max(|\hat{\mathbf{r}}_{pk}| - \zeta_{2} \mathbf{1}, \mathbf{0})||_{0} - T_{D}}{K_{max}-K},
\label{pfa_def}
\end{align}
where $K_{max} = \lceil N/2 \rceil$ is the maximum number of distinct peaks possible given an angular grid of length $N$. Defining the closest angle $\hat{\theta}_{k}^{*} \in \mathcal{S}^{2}_{\theta_{k}^{*}}$ to $\theta_{k}^{*}$ as
\begin{align}
	\hat{\theta}_{k}^{*} =\argmin \limits_{\theta \in \mathcal{S}^{2}_{\theta_{k}^{*}}}|\theta-\theta_{k}^{*}|,
\end{align}
and denoting the subset of indices that correspond to successful detections by $\mathcal{I}_{D} = \{ k \in {\mathbb{N}}: \mathds{1}(\mathcal{S}^{2}_{\theta_{k}^{*}}) \neq 0\}$, the RMSE is then given by 
\begin{align}
\textrm{RMSE} = \sqrt{\frac{1}{T_{D}} \sum_{k\in \mathcal{I}_{D}}(\hat{\theta}_{k}^{*}-\theta_{k}^{*})^{2}}.
\label{rmse_def}
\end{align}
In our experiments we fix $(\zeta_{1},\zeta_{2},\delta \theta) = (0.1, 5 \times10^{-3}, 2\Delta\theta)$.
\subsection{Experimental Setup}
\subsubsection{Array Imperfections}
\begin{figure}[t] 
	\centering
	\includegraphics[height=0.6cm,width=12cm]{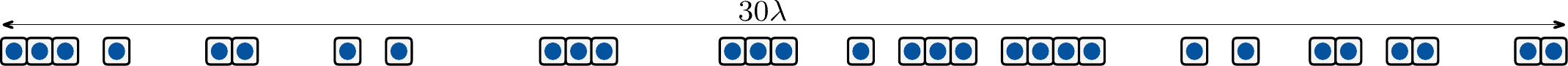}
	\caption{Topology of a 30-element sparse array subsampled from a ULA with an aperture of $30 \lambda$. The elements are positioned on a linear grid with $\lambda/2$ unit distance.}
    \label{array_topology}
\end{figure}

\begin{figure}[h!] 
	\centering
	\includegraphics[height=5cm,width=8.7cm]{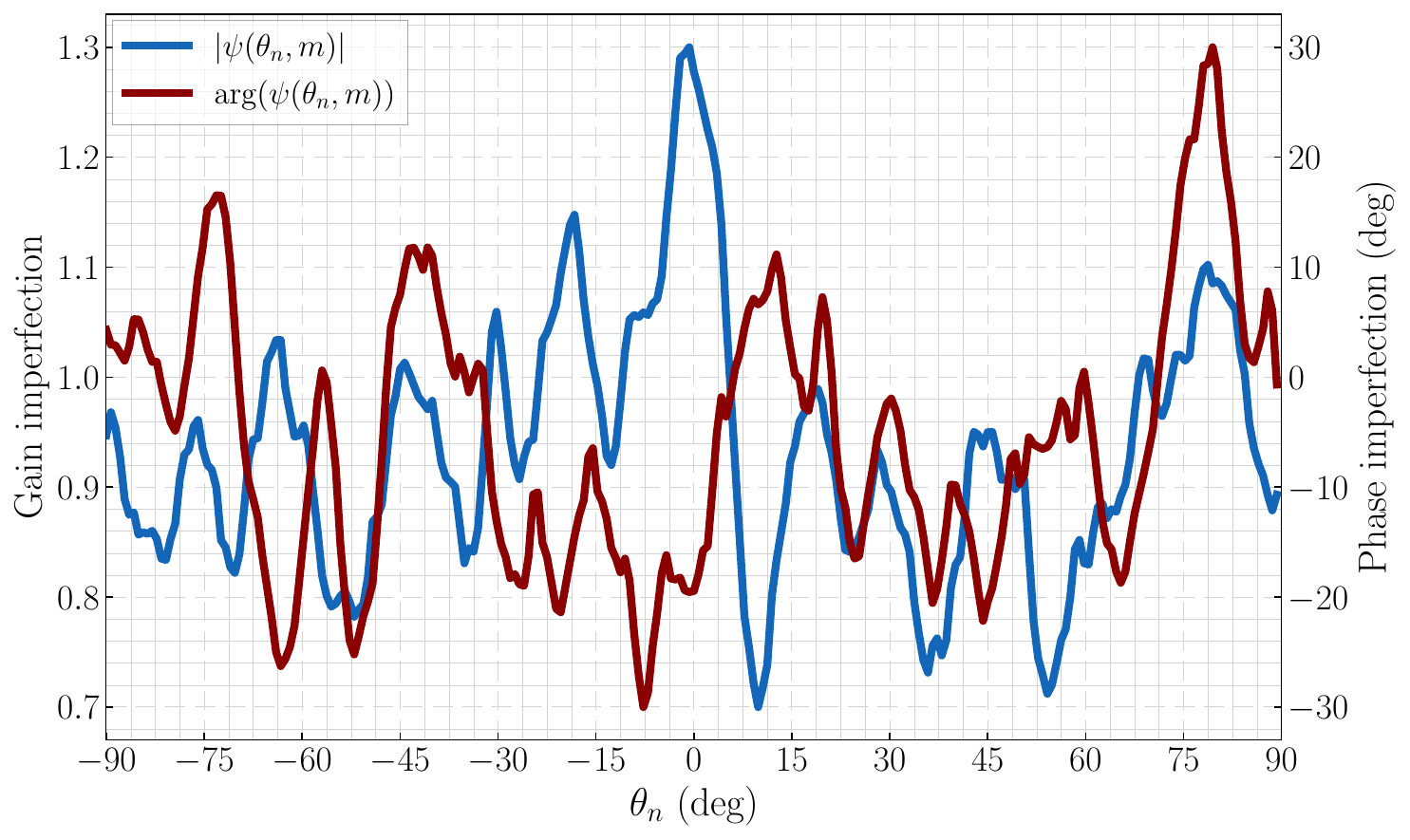}
	\caption{Gain-phase imperfection coefficient $\pmb{\psi}(\theta_{n_{1}},m)$ of element $m=6$ of the array as a function of the angular position $\theta_{n}$, with $\max(|\pmb{\psi}(\theta_{n_{1}},m)|)-\min(|\pmb{\psi}(\theta_{n_{1}},m)|)=0.6$ and  $\max(\arg(\pmb{\psi}(\theta_{n_{1}},m)))-\min(\arg(\pmb{\psi}(\theta_{n_{1}},m)))=60^{\circ}$}
    \label{gain_phase_imperfection_6}
\end{figure}

\begin{figure}[h!] 
	\centering
	\includegraphics[height=5cm,width=8.7cm]{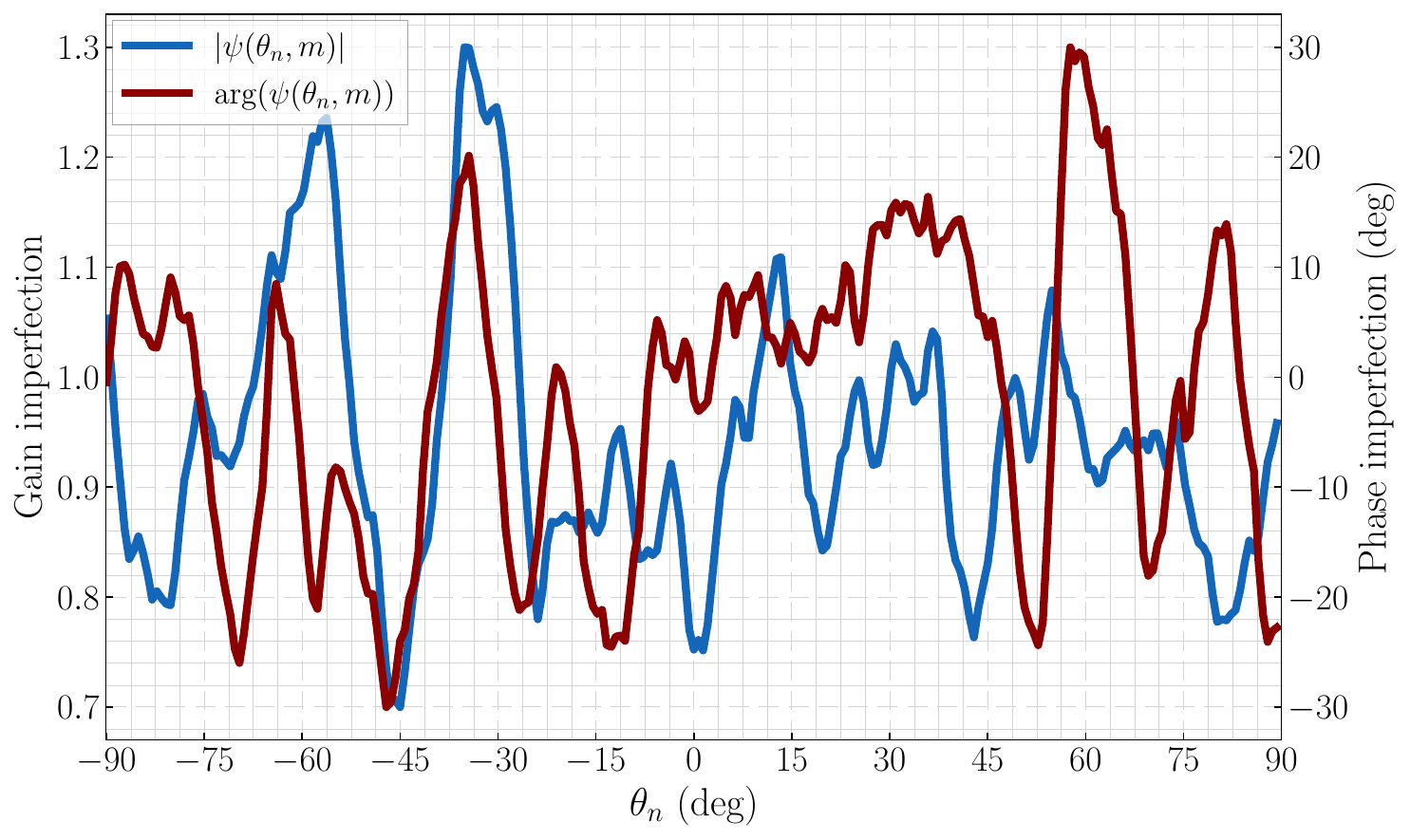}
	\caption{Gain-phase imperfection coefficient $\pmb{\psi}(\theta_{n_{1}},m)$ of element $m=17$ of the array.}
    \label{gain_phase_imperfection_17}
\end{figure}

We randomly subsample $M=30$ elements from a 60-element uniform linear array (ULA) of aperture width $30\lambda$, where the inter-element spacing is $\lambda/2$ as indicated in Fig. \ref{array_topology}. Next, we sample the array manifold over a uniform angular grid $\mathcal{G}_{\theta}$ of $N = 256$ points such that $\theta_{n} = \theta_{1} + \Delta \theta (n-1)$, for $n=1,\hdots,N$, $\theta_{1} = -90^{\circ}$ and $\Delta \theta = \frac{180}{256}^{\circ}$. For a given $m$-th sensor from the array, we create a model of an angular-dependent gain-phase set of imperfection coefficients $\{\pmb{\psi}(\theta_{n},m)\}_{n=1}^{N}$ by generating two Gaussian random vectors $\tilde{\mathbf{w}}_{1,m},\tilde{\mathbf{w}}_{2,m} \sim \mathcal{N}(\mathbf{0}, \mathbf{\Sigma}_{m})$ where $\mathbf{\Sigma}_{m} \in \mathbb{R}^{N\times N}$ is a Matérn 3/2 kernel \cite{matern} 
\begin{align}
\mathbf{\Sigma}_{m}(n_{1},n_{2}) =
\big(1 + \frac{ \sqrt{3}d({n_1},{n_2})}{\xi}\big) \exp \big( -\frac{\sqrt{3} d({n_1},{n_2})}{\xi} \big),
\end{align}
where $d({n_1},{n_2}) = |\theta_{n_{1}}-\theta_{n_2}|$, and $\xi>0$ controls the correlation between $\tilde{\mathbf{w}}_{i,m}(n_{1})$ and $\tilde{\mathbf{w}}_{i,m}(n_{2})$, for $i=1,2$, for a given separation $|\theta_{n_{1}}-\theta_{n_2}|$. Next we perform a scaling transformation on $\tilde{\mathbf{w}}_{i,m}$ as follows
\begin{align}
\mathbf{w}'_{i,m} = \frac{2V_{i}}{\max(\tilde{\mathbf{w}}_{i,m})-\min(\tilde{\mathbf{w}}_{i,m})}\tilde{\mathbf{w}}_{i,m},
\end{align}
where $V_{i}>0$. We can note that this mapping satisfies 
\begin{align}
\max(\mathbf{w}'_{i,m})-\min(\mathbf{w}'_{i,m}) = 2V_{i}.
\end{align} 
Finally, we apply the following set of transformations
\begin{align}
\mathbf{w}_{1,m} &= \mathbf{w}'_{1,m} + 1+V_{1}-\max(\mathbf{w}'_{1,m}) \\
\mathbf{w}_{2,m} &= \mathbf{w}'_{2,m} + V_{2}-\max(\mathbf{w}'_{2,m}).
\end{align}
It can be easily verified that the following holds
\begin{align}
\begin{cases}
\max(\mathbf{w}_{1,m}) = 1 + V_{1} \\
\min(\mathbf{w}_{1,m}) = 1 - V_{1}
\end{cases}
\quad 
\begin{cases}
\max(\mathbf{w}_{2,m}) =  V_{2} \\
\min(\mathbf{w}_{2,m}) = - V_{2}.
\end{cases}
\end{align}
We finally define the complex gain-phase imperfection coefficient $\pmb{\psi}(\theta_{n},m)$ at angle $\theta_{n}$ for the $m$-th array element as  
\begin{align}
\pmb{\psi}(\theta_{n},m) = \mathbf{w}_{1,m}(n) \exp \big(j \pi \frac{\mathbf{w}_{2,m}(n)}{180}\big).
\end{align}
 In our experiments, we fix $\xi = 5 ^{\circ}$, $V_{1} = 0.3$, and $V_{2}=30^{\circ}$ so that the gain imperfections for any sensor shows a range of $0.6$ whereas the phase imperfections show a variation range of $60^{\circ}$. Moreover the correlation between $\pmb{\psi}(\theta_{n_{1}},m)$ and $\pmb{\psi}(\theta_{n_{2}},m)$ begins to strongly decrease beyond an angular separation of $|\theta_{n_{1}}-\theta_{n_{2}}|>5 ^{\circ}$. Fig. \ref{gain_phase_imperfection_6} and Fig. \ref{gain_phase_imperfection_17} show the angular dependency obtained with these settings for the element $m=6$ and $m=17$, respectively.
\par
\begin{figure}[h!] 
	\centering
	\includegraphics[height=5cm,width=8cm]{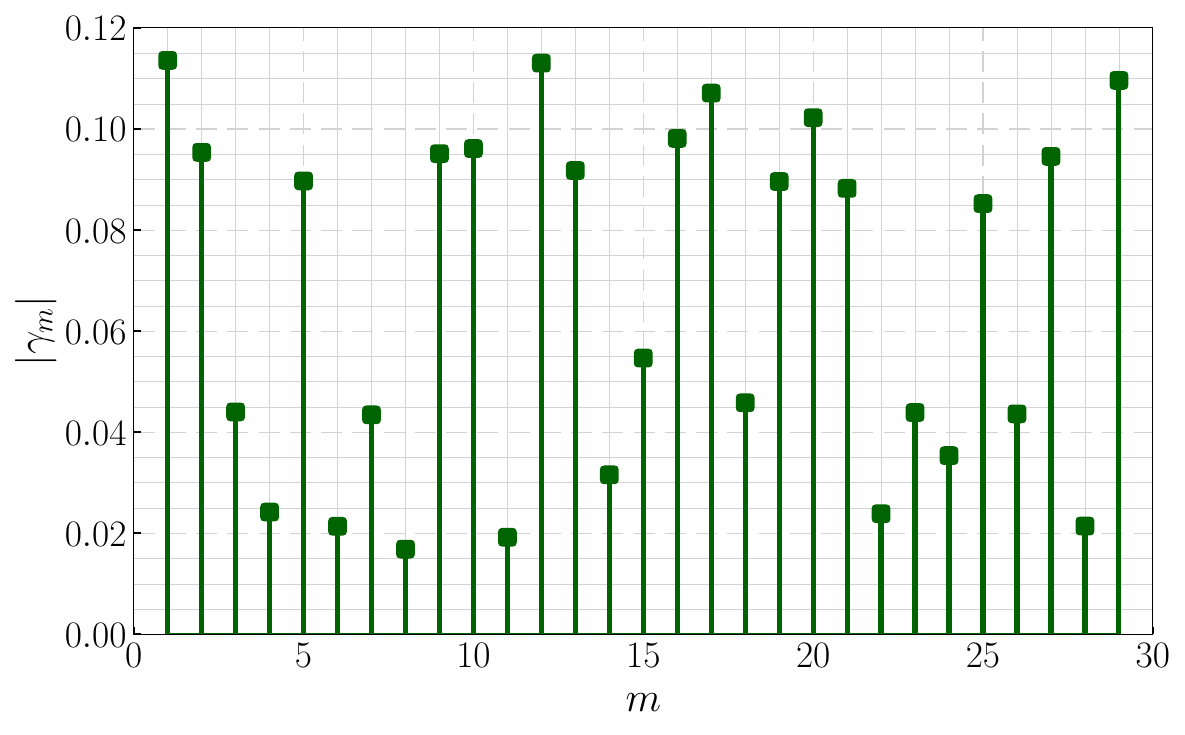}
	\caption{Amplitude of the mutual coupling coefficient $|\gamma_{m}|$ between element $m$ and $m+1$ for $m=1,\hdots M-1$. The mutual coupling strength decreases as the distance between two elements increases. }
    \label{mutual_coupling_coefficients}
\end{figure}

 Regarding the mutual coupling matrix $\mathbf{\Gamma}$ as defined in \eqref{mutual_coupling_matrix}, we generate a mutual coupling coefficient between element $m$ and $m+1$, for $1\le m < M$, as follows
\begin{align}
\tilde{\gamma}_{m} = \frac{A_{1}}{|p_{m+1}-p_{m}|}(1 + \sqrt{3}A_{2}u_{m}),
\end{align}
where  $p_{m}$ is the position of the $m$-th element $A_{1}>0$, $0<A_{2}<1/ \sqrt{3}$, and $u_{m}\sim \mathcal{U}(-1,1)$. We therefore have 
\begin{align}
\begin{cases}
\mathds{E}(\tilde{\gamma}_{m}) = \dfrac{A_{1}}{|p_{m+1}-p_{m}|} \\
\textrm{Var}(\tilde{\gamma}_{m}) = \big(A_{2} \mathds{E}(\tilde{\gamma}_{m})\big)^{2}.
\end{cases}
\end{align}
The final mutual coupling coefficient is then given by 
\begin{align}
\gamma_{m} = \tilde{\gamma}_{m} \exp(j \upsilon_{{m} }),     
\end{align}
where $\upsilon_{{m}} \sim \mathcal{U}(0, 2\pi)$. In our experiments, we set $(A_{1},A_{2}) = (0.1 \times \lambda/2, 0.1)$ so as to have an average coupling magnitude of $0.1$ when two elements are $\lambda/2$ apart. Additionally, when $|p_{m+1}-p_{m}| = \lambda/2$ we will have $\mathds{E}(|\gamma_{m}|) = 0.1$, $\max(|\gamma_{m}|) \approx 0.12$ and $\min(|\gamma_{m}|) \approx 0.08$. Fig. \ref{mutual_coupling_coefficients} shows the resulting amplitudes of the mutual coupling coefficients $|\gamma_{m}|$ for the elements of the array in Fig. \ref{array_topology}. The mutual coherence of the generated dictionary is measured to be $\mu(\mathbf{D})=0.9986$, which falls under the highly-coherent regime.
\subsubsection{Dataset Generation}
\par To measure the recovery performance of the proposed algorithms we create a dataset according to the following procedure. We first set the number of sources $K=10$ and generate $K$ amplitudes $\{r_{k}\}_{k=1}^{K}$ where $|r_{k}| \sim \mathcal{U}(0.05,1)$ and $\textrm{arg}(r_{k}) \sim \mathcal{U}(0,2\pi)$ along with $K$ random normalized frequencies $\{f_{k}^{*}\}_{k=1}^{K}$ from $[-0.5,0.5[$ such that the minimum separation between any two frequencies is at least $1/(3M)$, since the normalized frequencies separation determines the resolvability of sources \cite{super-resolution}. Afterwards, we convert the normalized frequencies $\{f_{k}^{*}\}_{k=1}^{K}$ to their corresponding angles $\{\theta_{k}^{*}\}_{k=1}^{K}$ using the mapping $\theta_{k}^{*} =  \sin^{-1}(2f_{k}^{*})$. Next, since $\theta_{k}^{*} \in [-90^{\circ}, 90^{\circ}]$ is a continuous variable whose value may not necessarily lie exactly on the angular grid $\mathcal{G}_{\theta}$, we first find the two closest angles $\theta_{k,1}, \theta_{k,2} \in \mathcal{G}_{\theta}$ such that $\theta_{k,1}<\theta_{k}^{*}<\theta_{k,2}$ and we then generate $\boldsymbol{\psi}(\theta_{k}^{*})$ by performing a simple linear interpolation between $\boldsymbol{\psi}(\theta_{k,1})$ and $\boldsymbol{\psi}(\theta_{k,2})$ as follows
\begin{align}
\boldsymbol{\psi}(\theta_{k}^{*}) = \boldsymbol{\psi}(\theta_{k,1}) + \frac{\theta_{k}^{*}-\theta_{k,1}}{\theta_{k,2}-\theta_{k,1}}(\boldsymbol{\psi}(\theta_{k,2})-  \boldsymbol{\psi}(\theta_{k,1})).
\end{align}
The final array steering vector $\mathbf{a}(\theta_{k}^{*})$ is then evaluated directly using \eqref{real_manifold}. The same procedure is repeated for the $K$ sources, and the final measurement vector is then given by \eqref{signal_model}. We evaluate the performance using the metrics defined in \ref{metrics_definition} by varying the noise variance as $\sigma^{2}_{l} = 10^{-l/2}$ with $l=0, 1, \hdots, 8$ and computing the average metric value over $10^{3}$ different measurement vectors for each variance level $\sigma^{2}_{l}$.
\subsection{Results}
We compare the performance of $SR\text{-}TL_{1}$ against the LASSO, square-root LASSO, the difference of $L_{1}$ and $L_{2}$ norms ($L_{1}L_{2}$-norm) \cite{l1l2_norm}, $TL_{1}$-norm, the Minmax Concave Penalty (MCP) \cite{mcp}, and SBL. We plot the $P_{d}$, $P_{fa}$, and RMSE curves of each framework as a function of the noise variance $\sigma^{2}$ while also varying the regularizer $\kappa$ for each $\sigma^{2}$. We fix the number of iterations for SBL to $10^{3}$. For the $L_{1}L_{2}$, $TL_{1}$, MCP, and $SR\text{-}TL_{1}$ frameworks we use the DCA algorithm with $t_{max} = 50$ outer iterations and $l_{max} = 20$ inner iterations for the ADMM solution of the subproblem, resulting in $10^{3}$ effective iterations. We solve the LASSO and square-root LASSO using ADMM as well and we similarly fix the number of iterations to $10^{3}$. For both $TL_{1}$ and $SR\text{-}TL_{1}$ we set $\alpha = 1$, and we additionally set $(\rho_{1}, \rho_{2}) = (1, 10)$ for $SR\text{-}TL_{1}$. Finally, for MCP we set $\gamma = 0.5$.  We initialize all the vectors with the null vector.

\begin{sidewaysfigure}[htbp]
    \centering
    \begin{subfigure}[t]{0.22\textwidth}
        \centering
        \includegraphics[width=1.1\linewidth, height = 0.9\linewidth]{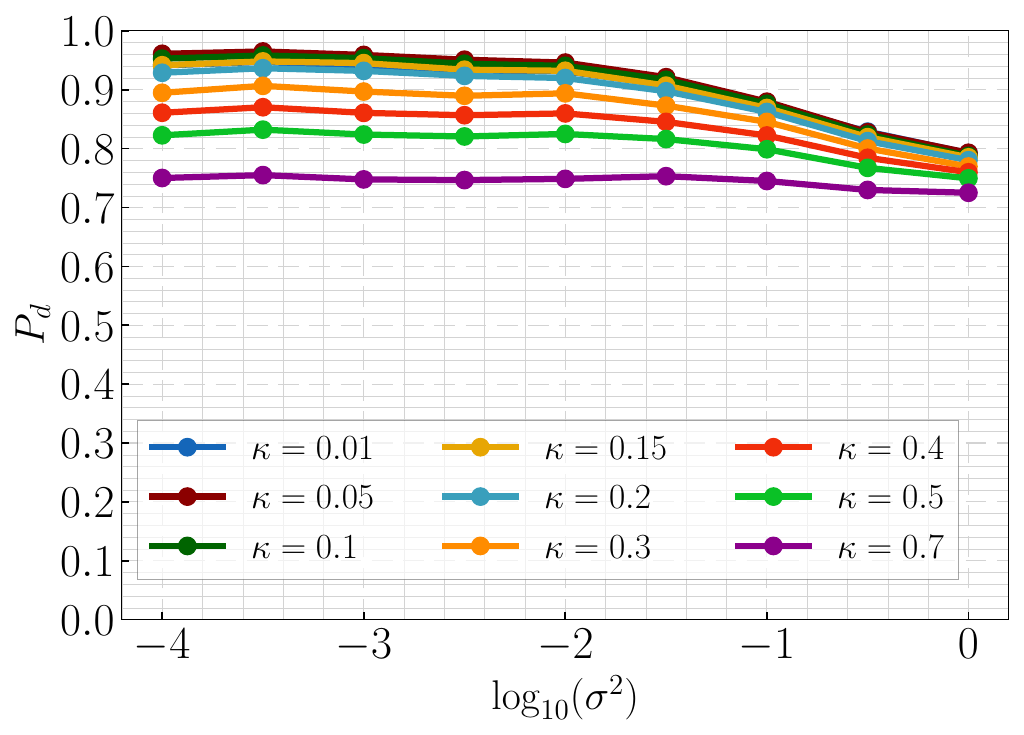}
        \caption{LASSO}
        \label{lasso_detection_rate}
    \end{subfigure}
    \hfil
    \begin{subfigure}[t]{0.22\textwidth}
        \centering
        \includegraphics[width=1.1\linewidth, height = 0.9\linewidth]{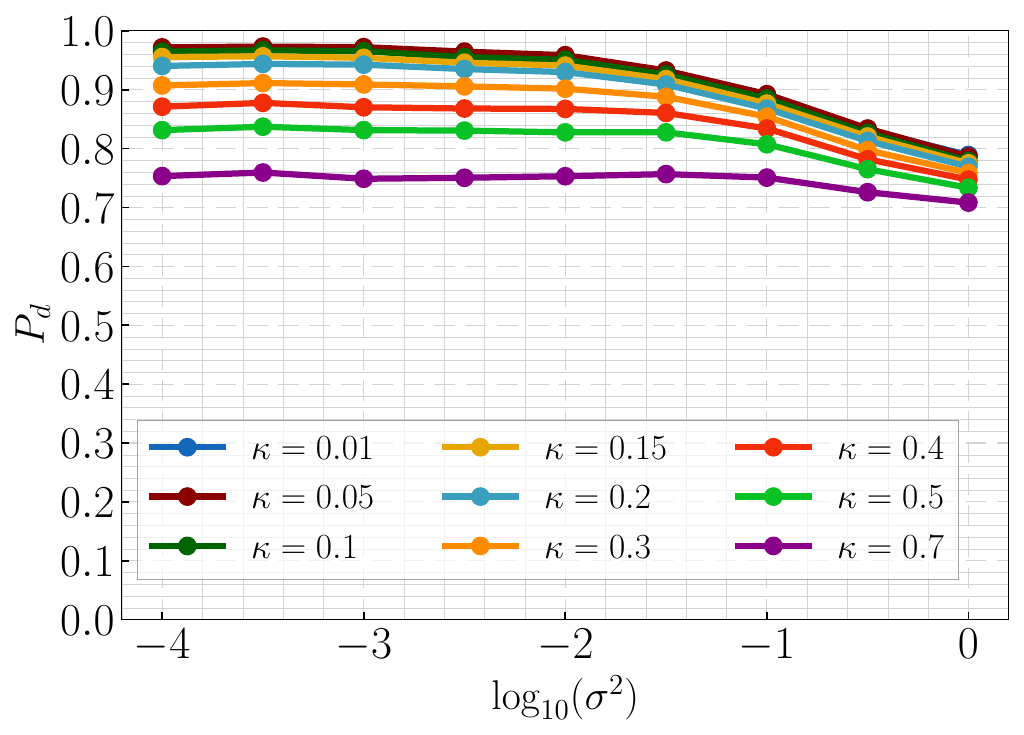}
        \caption{$L_{1}L_{2}$}
        \label{l1l2_detection_rate}
    \end{subfigure}
    \hfil
    \begin{subfigure}[t]{0.22\textwidth}
        \centering
        \includegraphics[width=1.1\linewidth, height = 0.9\linewidth]{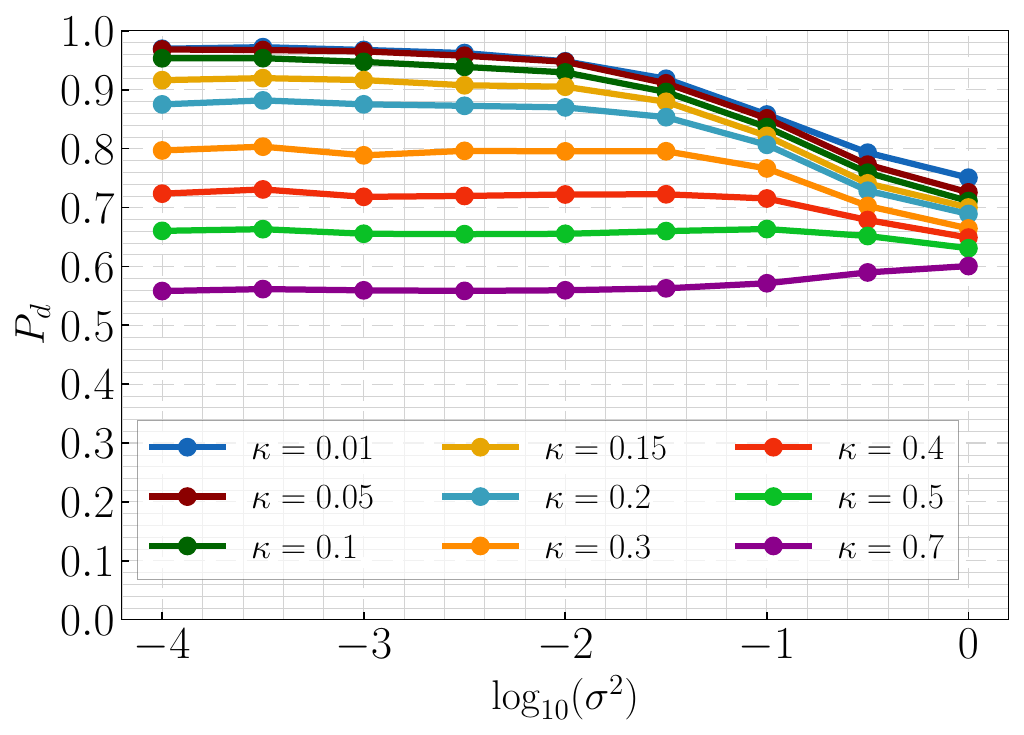}
        \caption{$TL_{1}$}
        \label{tl1_detection_rate}
    \end{subfigure}
    \hfil
    \begin{subfigure}[t]{0.22\textwidth}
        \centering
        \includegraphics[width=1.1\linewidth, height = 0.9\linewidth]{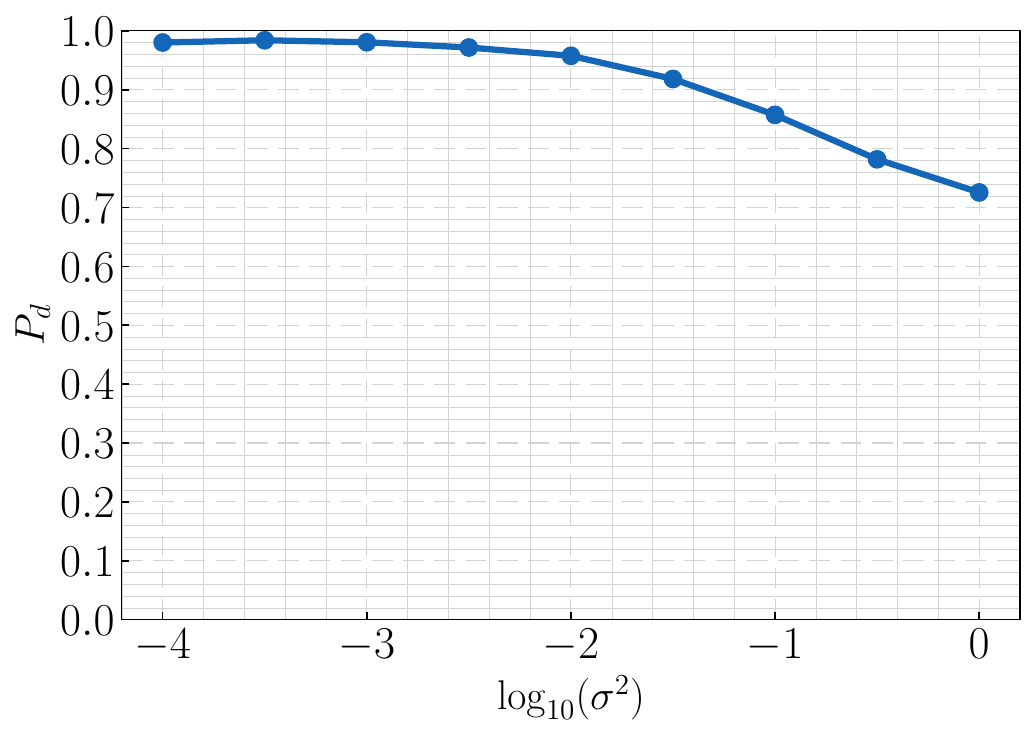}
        \caption{SBL}
        \label{sbl_detection_rate}
    \end{subfigure}
    \hfil
    \begin{subfigure}[t]{0.22\textwidth}
        \centering
        \includegraphics[width=1.1\linewidth, height = 0.9\linewidth]{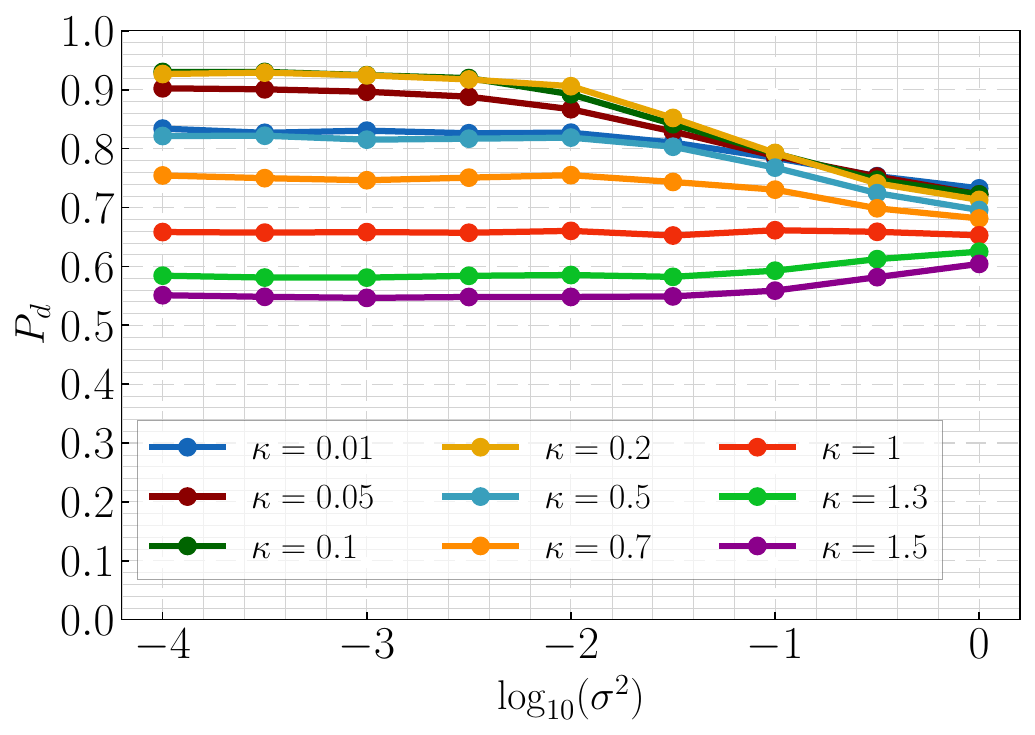}
        \caption{MCP}
        \label{mcp_detection_rate}
    \end{subfigure}
    \hspace{0.3cm}
    \begin{subfigure}[t]{0.22\textwidth}
        \centering
        \includegraphics[width=1.1\linewidth, height = 0.9\linewidth]{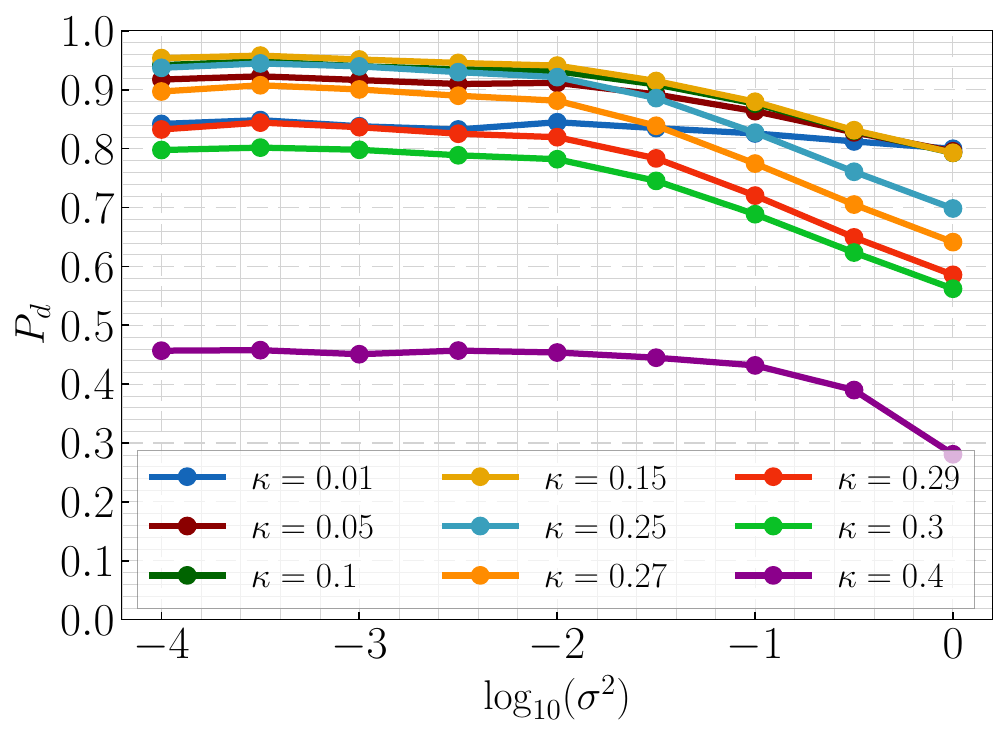}
        \caption{square-root LASSO}
        \label{sr_lasso_detection_rate}
    \end{subfigure}
    \hspace{0.3cm}
    \begin{subfigure}[t]{0.22\textwidth}
        \centering
        \includegraphics[width=1.1\linewidth, height = 0.9\linewidth]{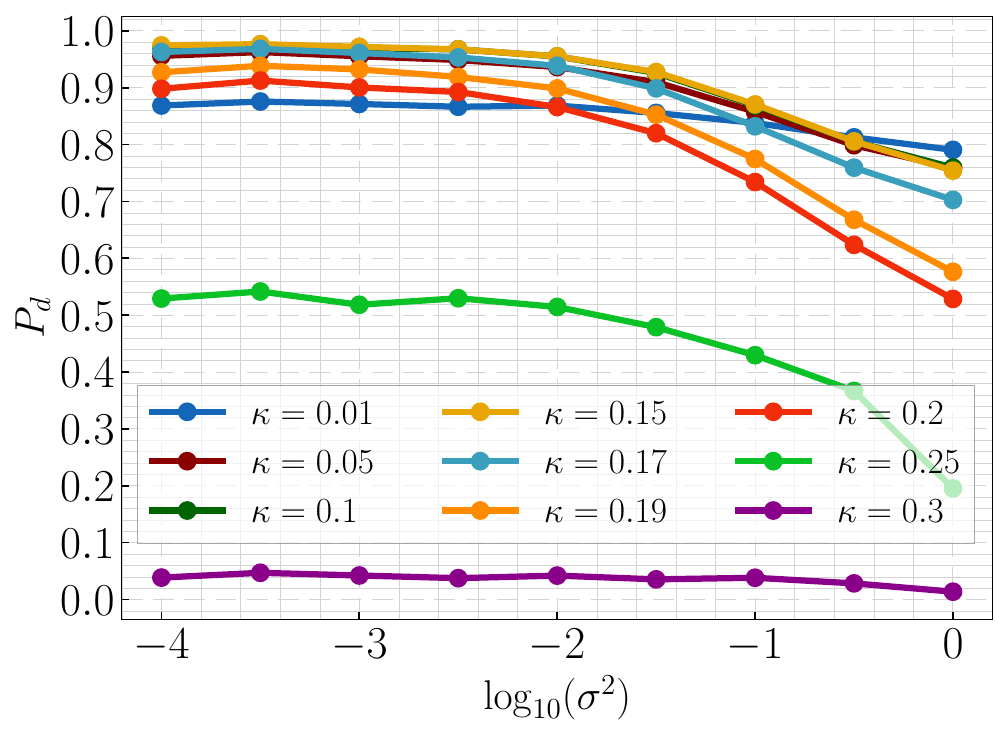}
        \caption{$SR\text{-}TL_{1}$}
        \label{sr_tl1_detection_rate}
    \end{subfigure}
    \caption{Detection rate performance $P_{d}$ as a function of $\sigma^{2}$ and the regularizer $\kappa$ for (a) LASSO, (b) $L_{1}L_{2}$, (c) $TL_{1}$, (d) SBL, (e) MCP, (f) square-root LASSO, and (g) $SR\text{-}TL_{1}$.}

\end{sidewaysfigure}
\begin{sidewaysfigure}[htbp]
    \centering
    \begin{subfigure}[t]{0.22\textwidth}
        \centering
        \includegraphics[width=1.1\linewidth, height = 0.9\linewidth]{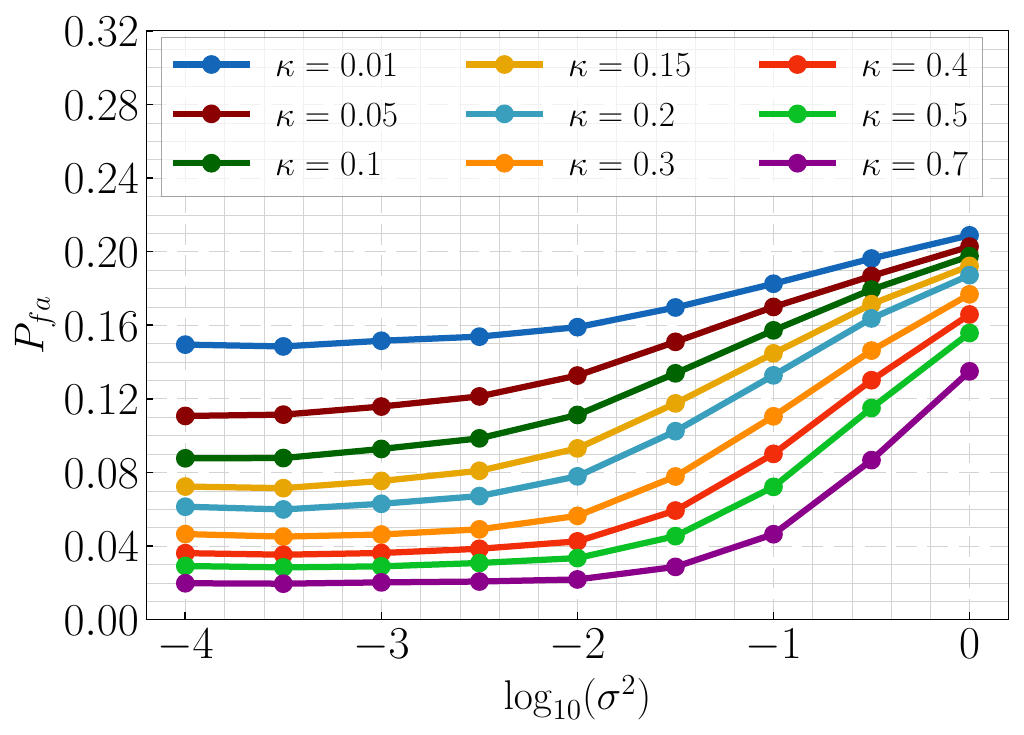}
        \caption{LASSO}
        \label{lasso_pfa}
    \end{subfigure}
    \hfil
    \begin{subfigure}[t]{0.22\textwidth}
        \centering
        \includegraphics[width=1.1\linewidth, height = 0.9\linewidth]{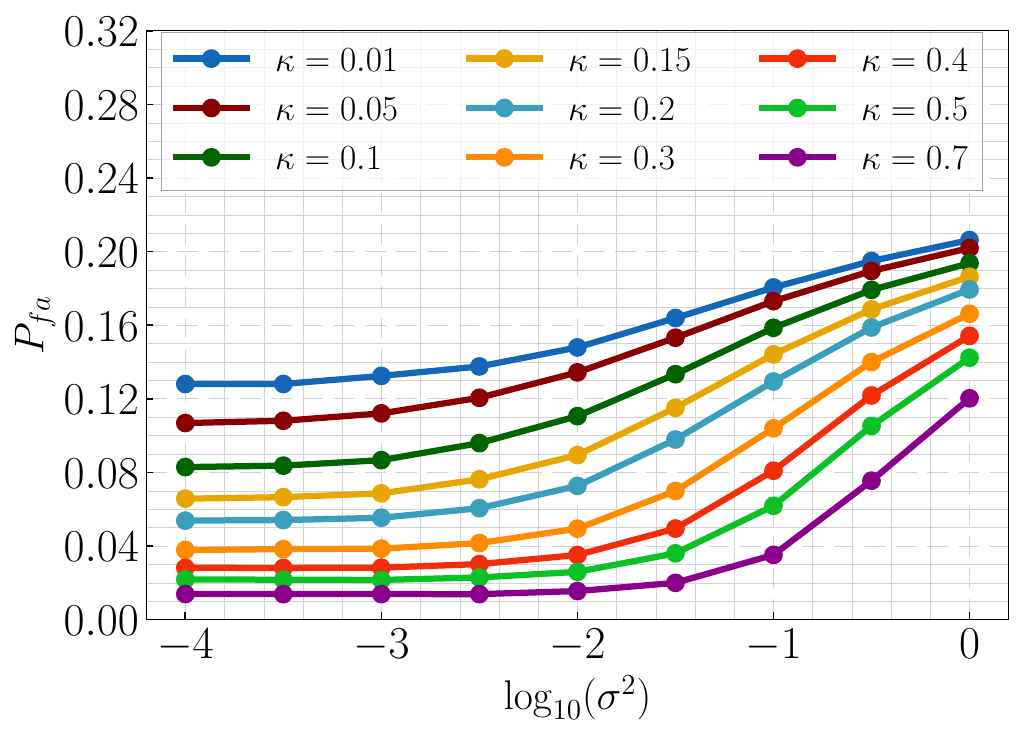}
        \caption{$L_{1}L_{2}$}
        \label{l1l2_pfa}
    \end{subfigure}
    \hfil
    \begin{subfigure}[t]{0.22\textwidth}
        \centering
        \includegraphics[width=1.1\linewidth, height = 0.9\linewidth]{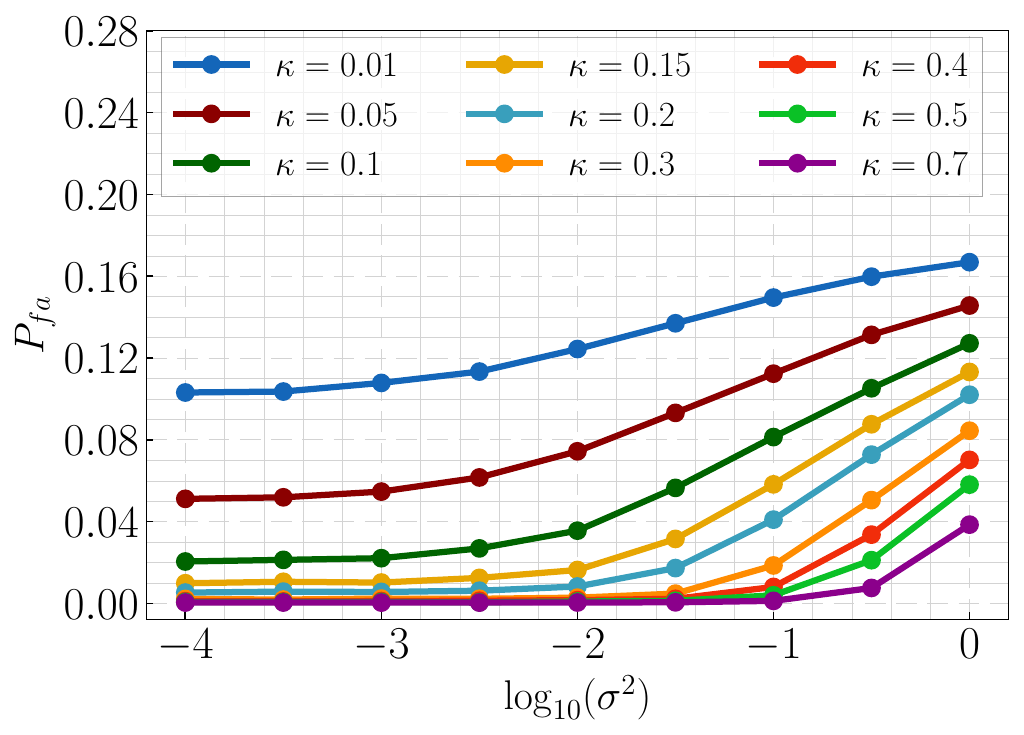}
        \caption{$TL_{1}$}
        \label{tl1_pfa}
    \end{subfigure}
    \hfil
    \begin{subfigure}[t]{0.22\textwidth}
        \centering
        \includegraphics[width=1.1\linewidth, height = 0.9\linewidth]{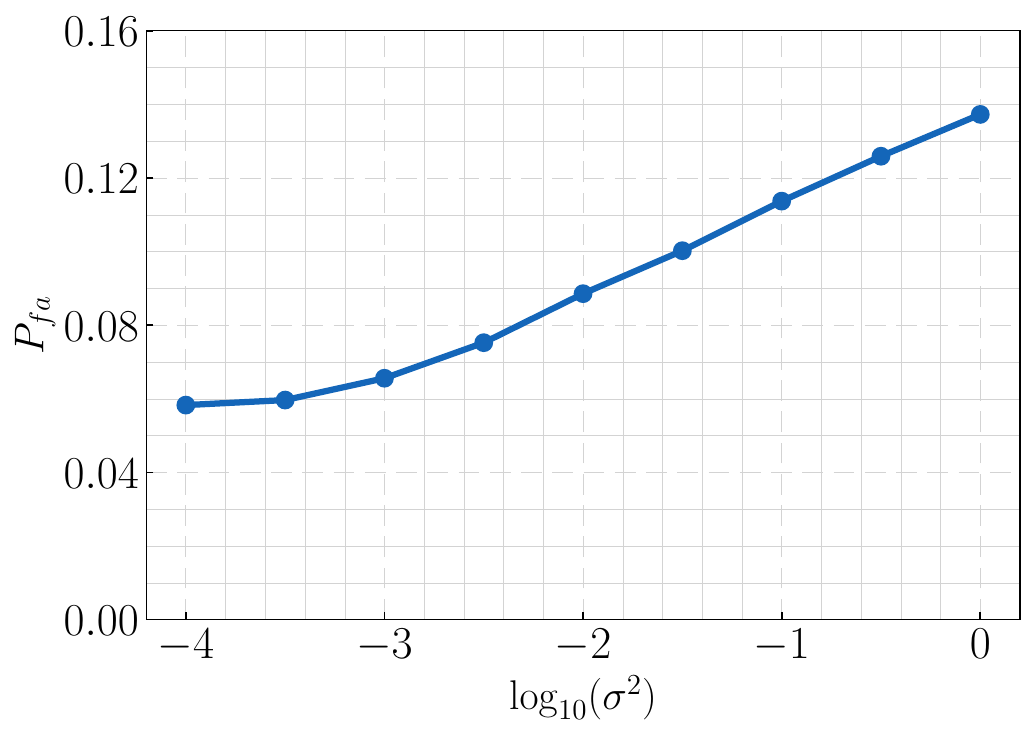}
        \caption{SBL}
        \label{sbl_pfa}
    \end{subfigure}
    \hfil
    \begin{subfigure}[t]{0.22\textwidth}
        \centering
        \includegraphics[width=1.1\linewidth, height = 0.9\linewidth]{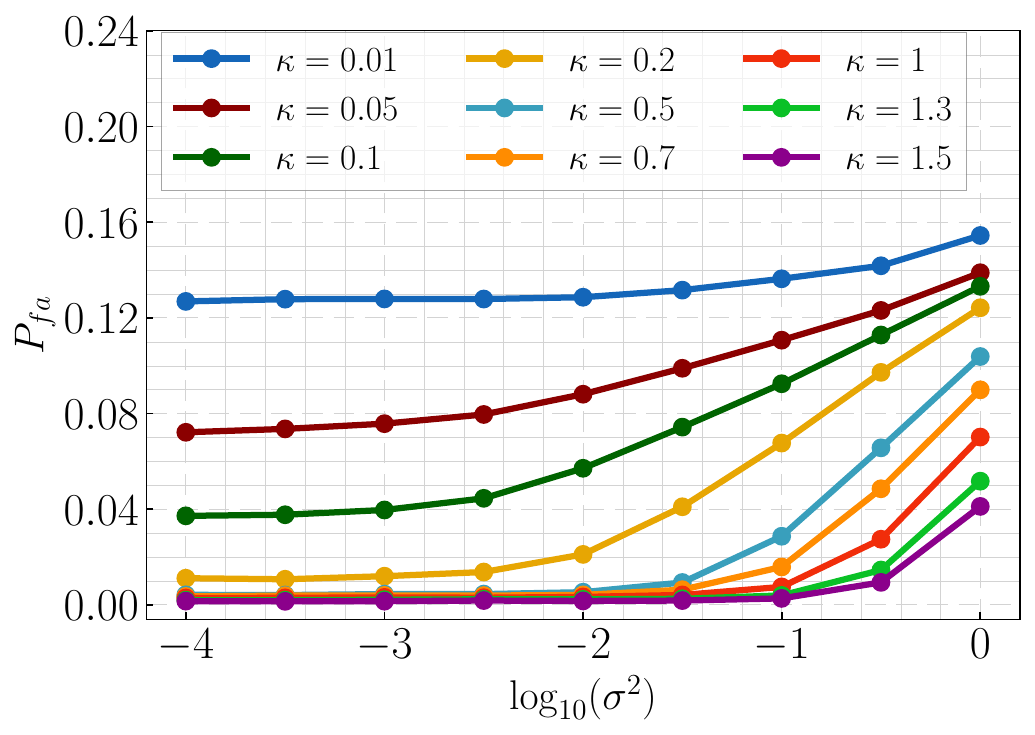}
        \caption{MCP}
        \label{mcp_pfa}
    \end{subfigure}
    \hspace{0.3cm}
    \begin{subfigure}[t]{0.22\textwidth}
        \centering
        \includegraphics[width=1.1\linewidth, height = 0.9\linewidth]{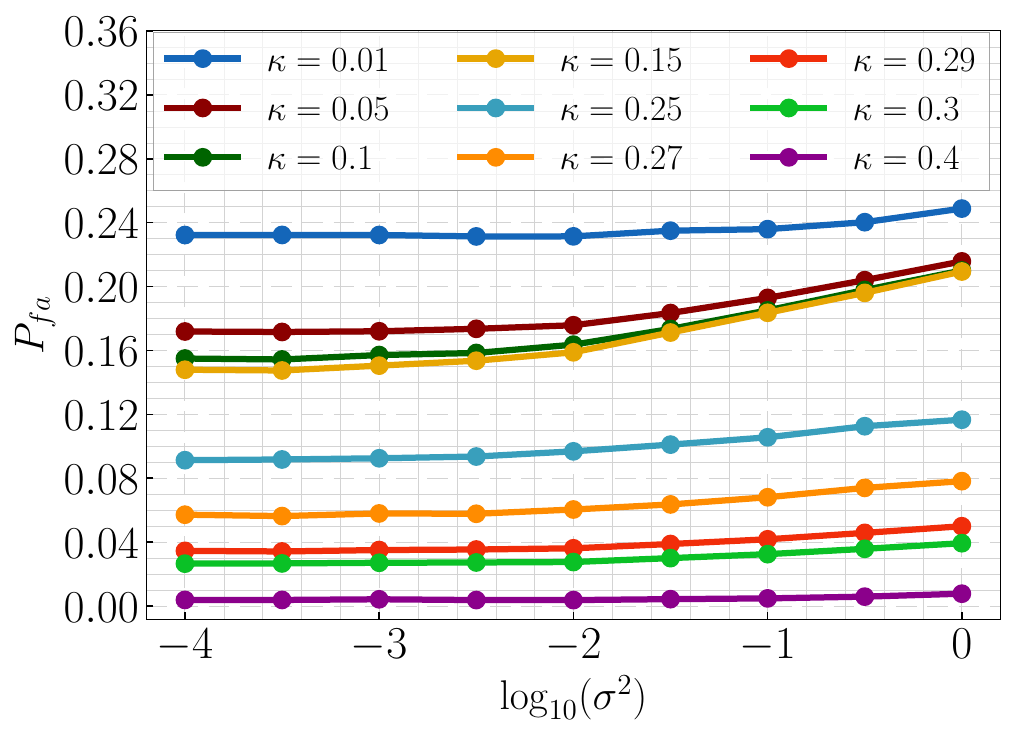}
        \caption{square-root LASSO}
        \label{sr_lasso_pfa}
    \end{subfigure}
    \hspace{0.3cm}
    \begin{subfigure}[t]{0.22\textwidth}
        \centering
        \includegraphics[width=1.1\linewidth, height = 0.9\linewidth]{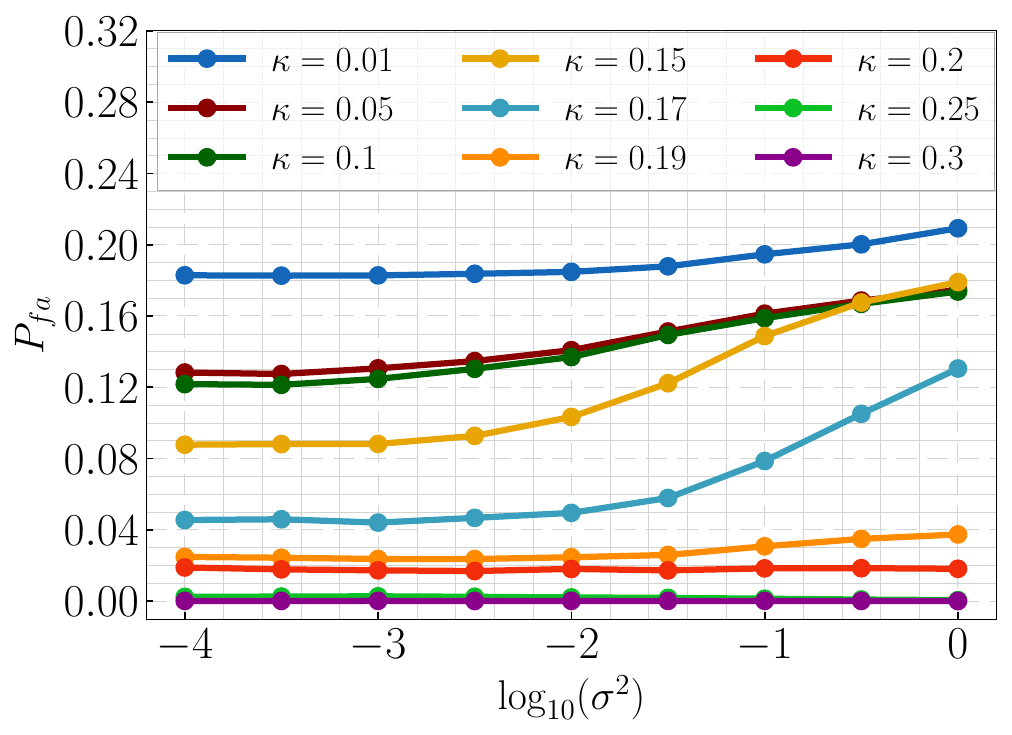}
        \caption{$SR\text{-}TL_{1}$}
        \label{sr_tl1_pfa}
    \end{subfigure}
    \caption{False alarm rate $P_{fa}$ as a function of $\sigma^{2}$ and the regularizer $\kappa$ for (a) LASSO, (b) $L_{1}L_{2}$, (c) $TL_{1}$, (d) SBL, (e) MCP, (f) square-root LASSO, and (g) $SR\text{-}TL_{1}$.}
\end{sidewaysfigure}
\begin{sidewaysfigure}[htbp]
    \centering
    \begin{subfigure}[t]{0.22\textwidth}
        \centering
        \includegraphics[width=1.1\linewidth, height = 0.9\linewidth]{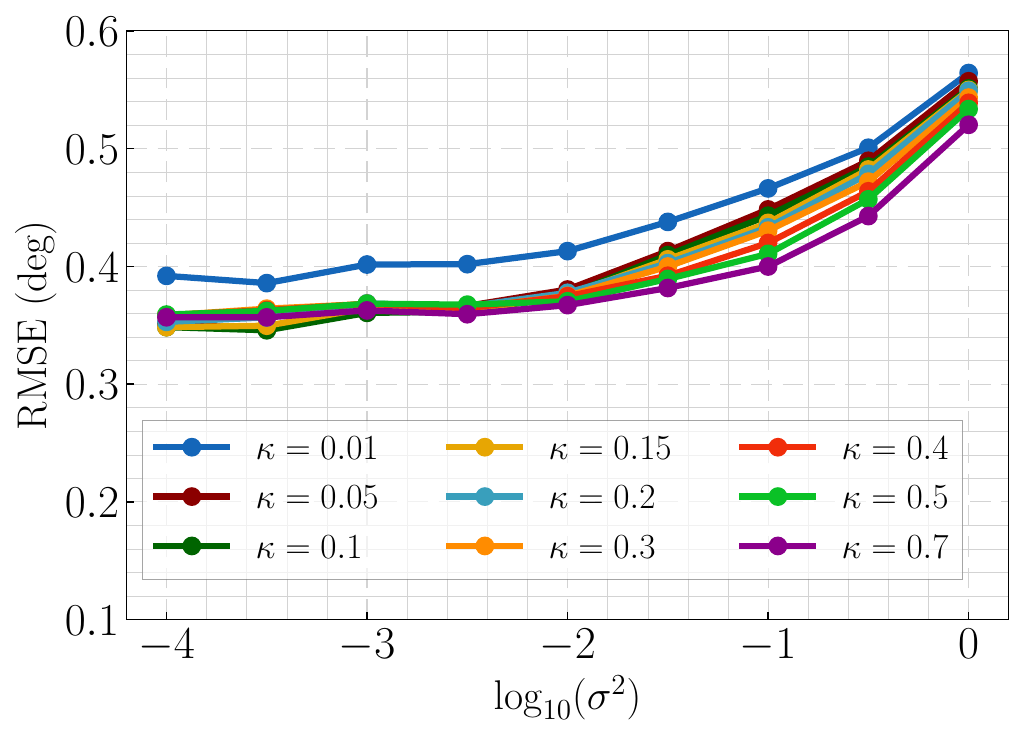}
        \caption{LASSO}
        \label{lasso_rmse}
    \end{subfigure}
    \hfil
    \begin{subfigure}[t]{0.22\textwidth}
        \centering
        \includegraphics[width=1.1\linewidth, height = 0.9\linewidth]{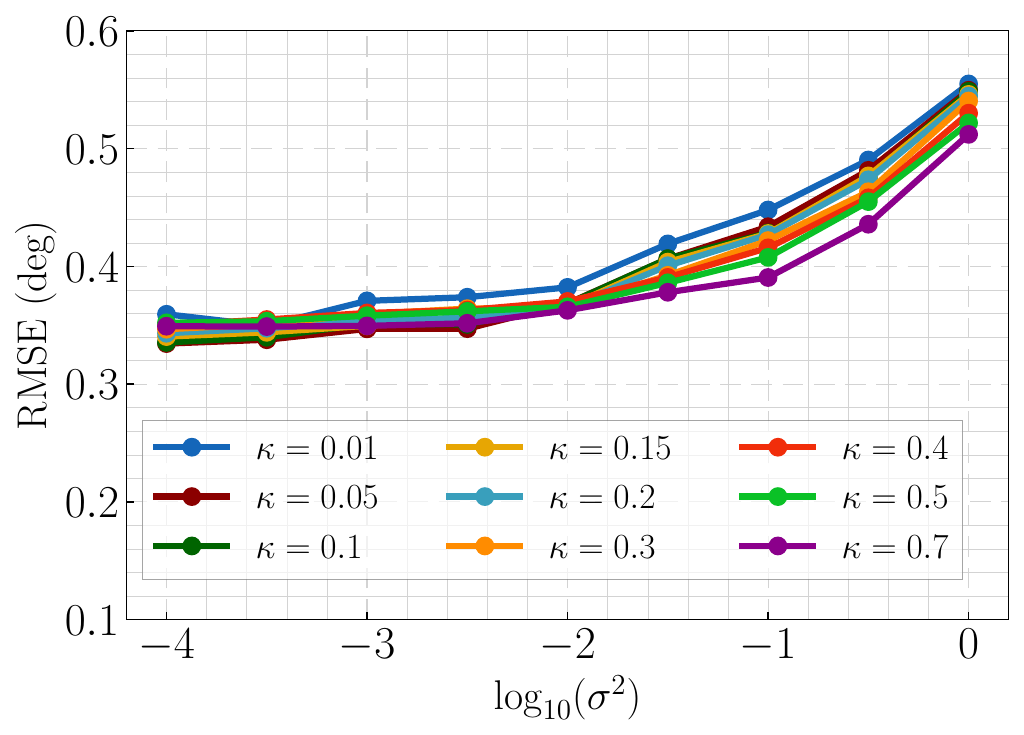}
        \caption{$L_{1}L_{2}$}
        \label{l1l2_rmse}
    \end{subfigure}
    \hfil
    \begin{subfigure}[t]{0.22\textwidth}
        \centering
        \includegraphics[width=1.1\linewidth, height = 0.9\linewidth]{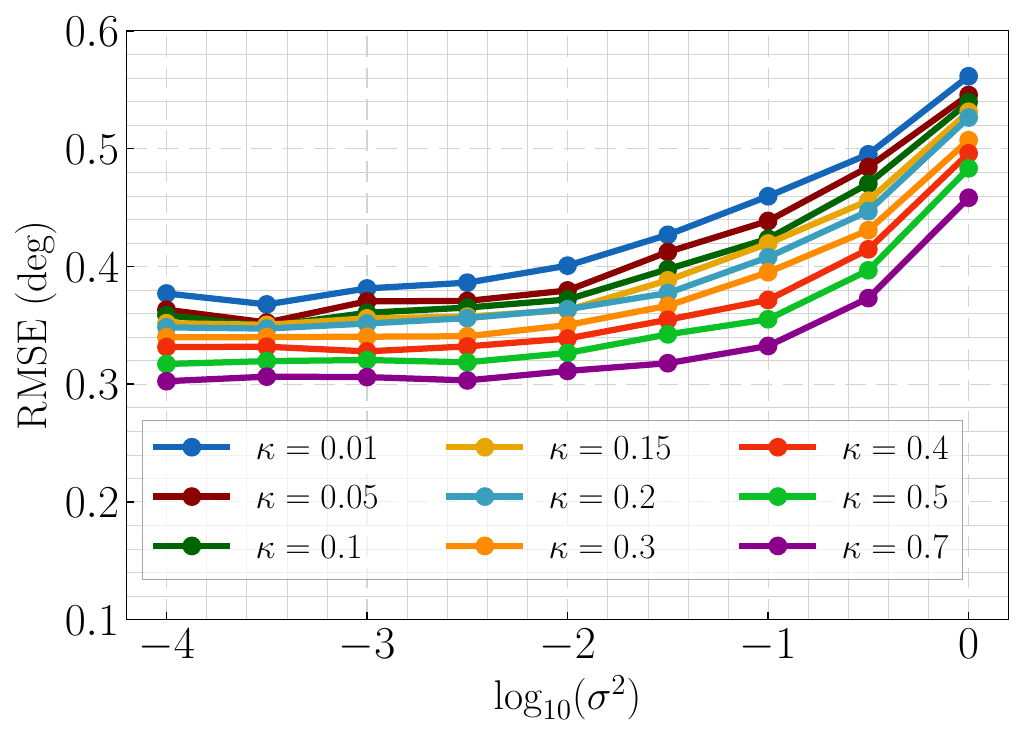}
        \caption{$TL_{1}$}
        \label{tl1_rmse}
    \end{subfigure}
    \hfil
    \begin{subfigure}[t]{0.22\textwidth}
        \centering
        \includegraphics[width=1.1\linewidth, height = 0.9\linewidth]{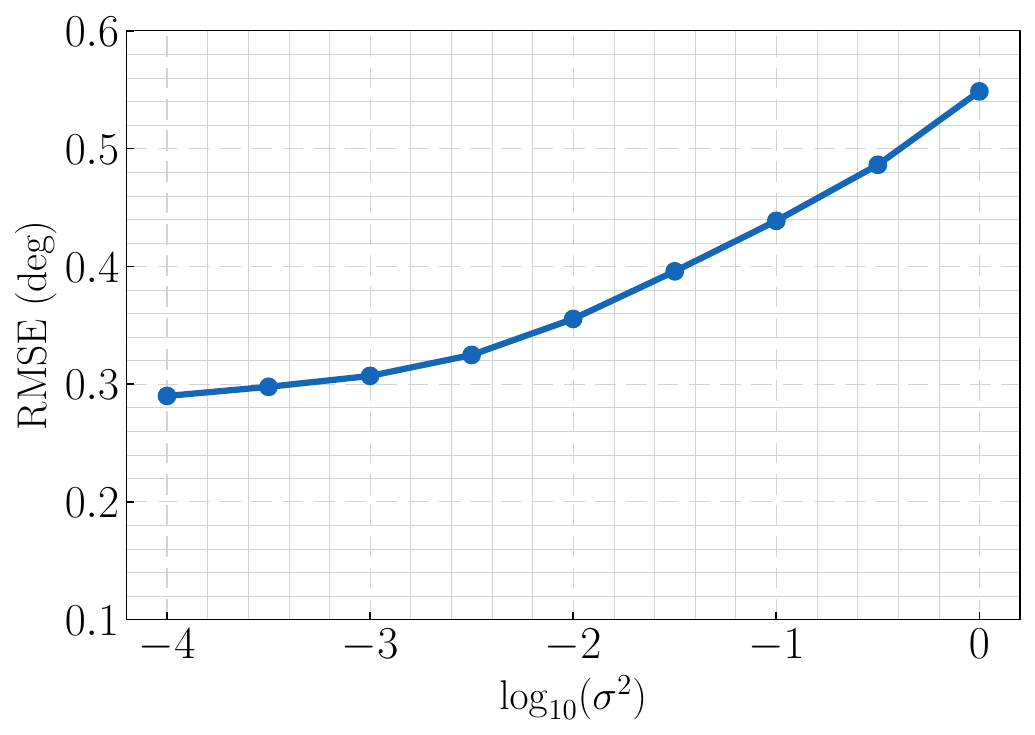}
        \caption{SBL}
        \label{sbl_rmse}
    \end{subfigure}
    \hfil
    \begin{subfigure}[t]{0.22\textwidth}
        \centering
        \includegraphics[width=1.1\linewidth, height = 0.9\linewidth]{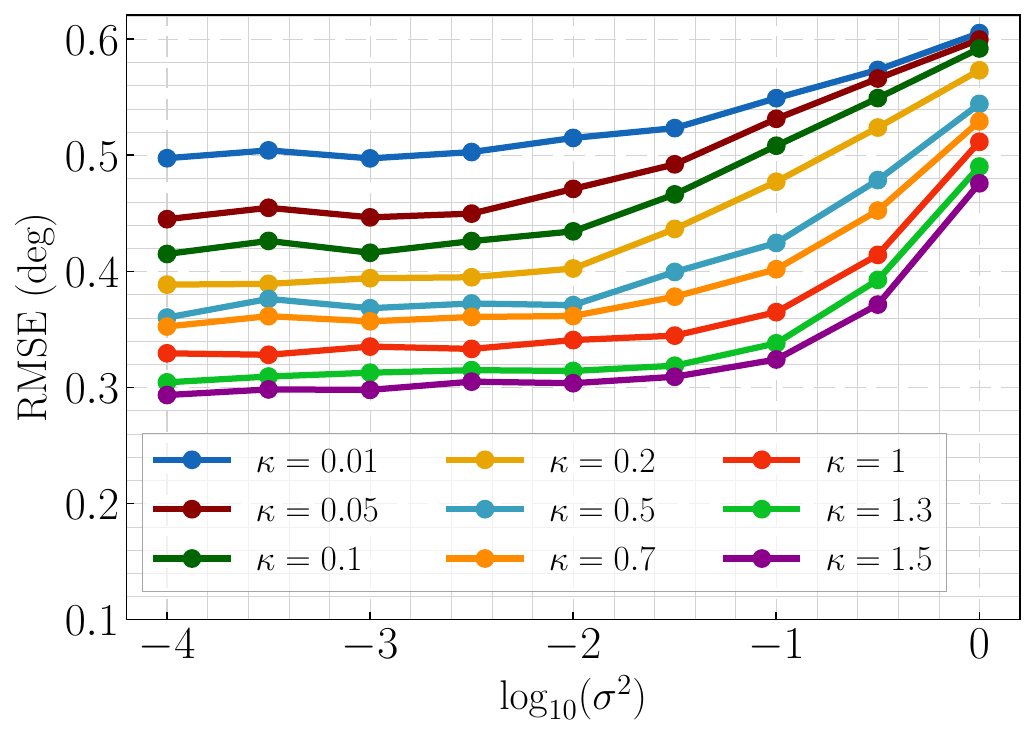}
        \caption{MCP}
        \label{mcp_rmse}
    \end{subfigure}
    \hspace{0.3cm}
    \begin{subfigure}[t]{0.22\textwidth}
        \centering
        \includegraphics[width=1.1\linewidth, height = 0.9\linewidth]{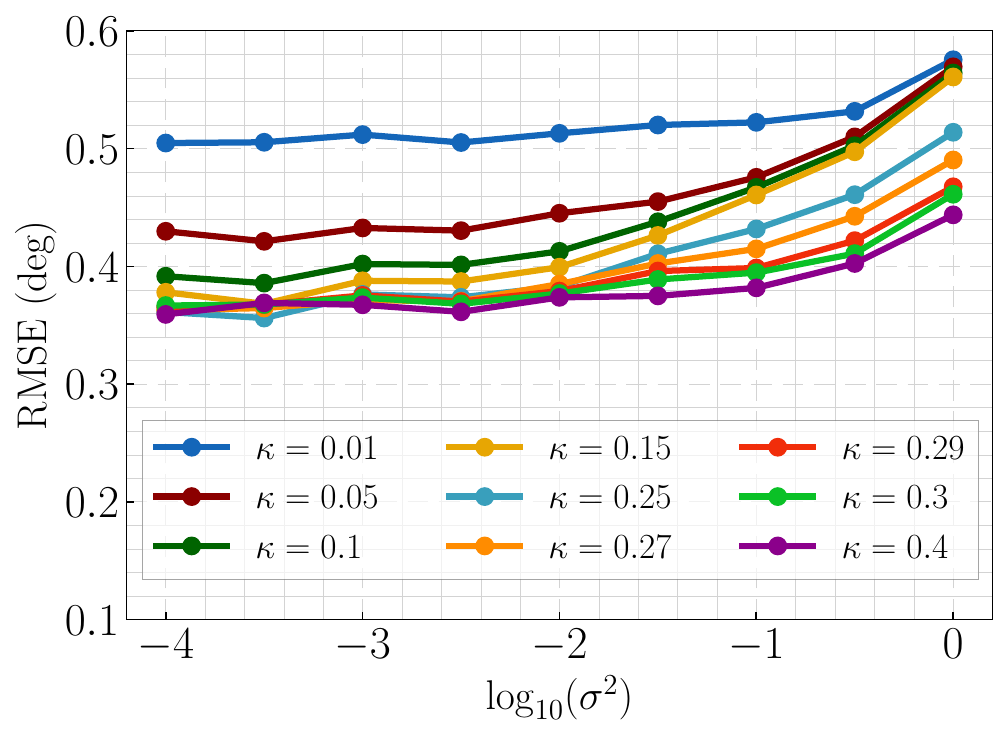}
        \caption{square-root LASSO}
        \label{sr_lasso_rmse}
    \end{subfigure}
    \hspace{0.3cm}
    \begin{subfigure}[t]{0.22\textwidth}
        \centering
        \includegraphics[width=1.1\linewidth, height = 0.9\linewidth]{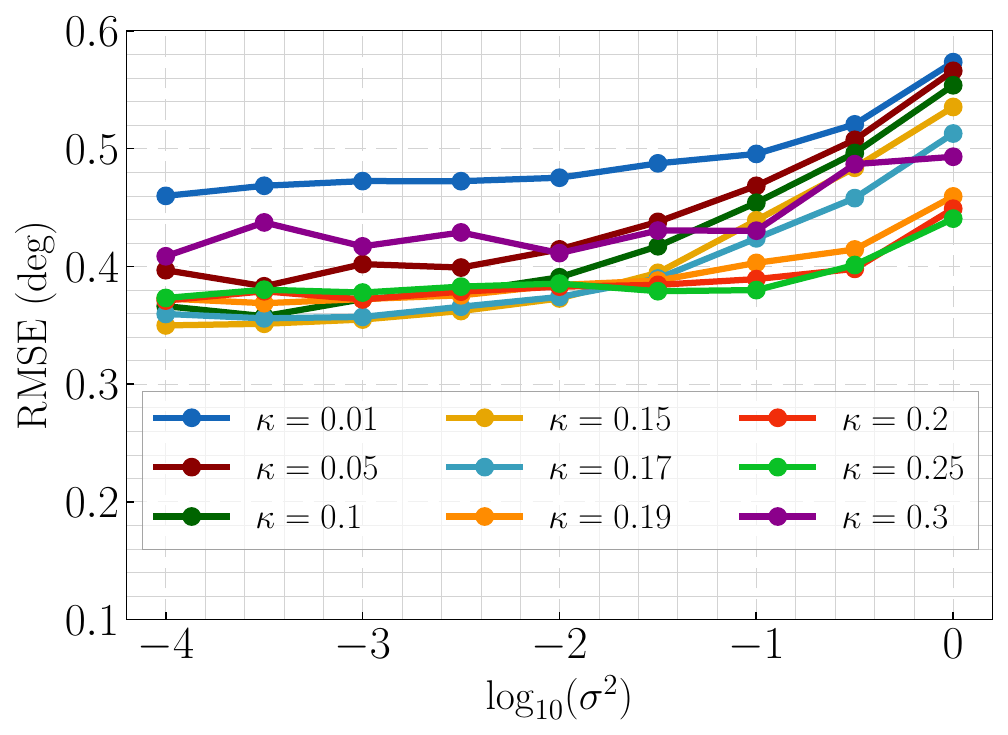}
        \caption{$SR\text{-}TL_{1}$}
        \label{sr_tl1_rmse}
    \end{subfigure}
    \caption{RMSE as a function of $\sigma^{2}$ and the regularizer $\kappa$ for (a) LASSO, (b) $L_{1}L_{2}$, (c) $TL_{1}$, (d) SBL, (e) MCP, (f) square-root LASSO, and (g) $SR\text{-}TL_{1}$.}
\end{sidewaysfigure}

\begin{figure}[!p]
    \centering
    \begin{subfigure}{\textwidth}
        \centering
        \makebox[\textwidth][c]{%
          \includegraphics[width=16cm, height=7cm, keepaspectratio=false]{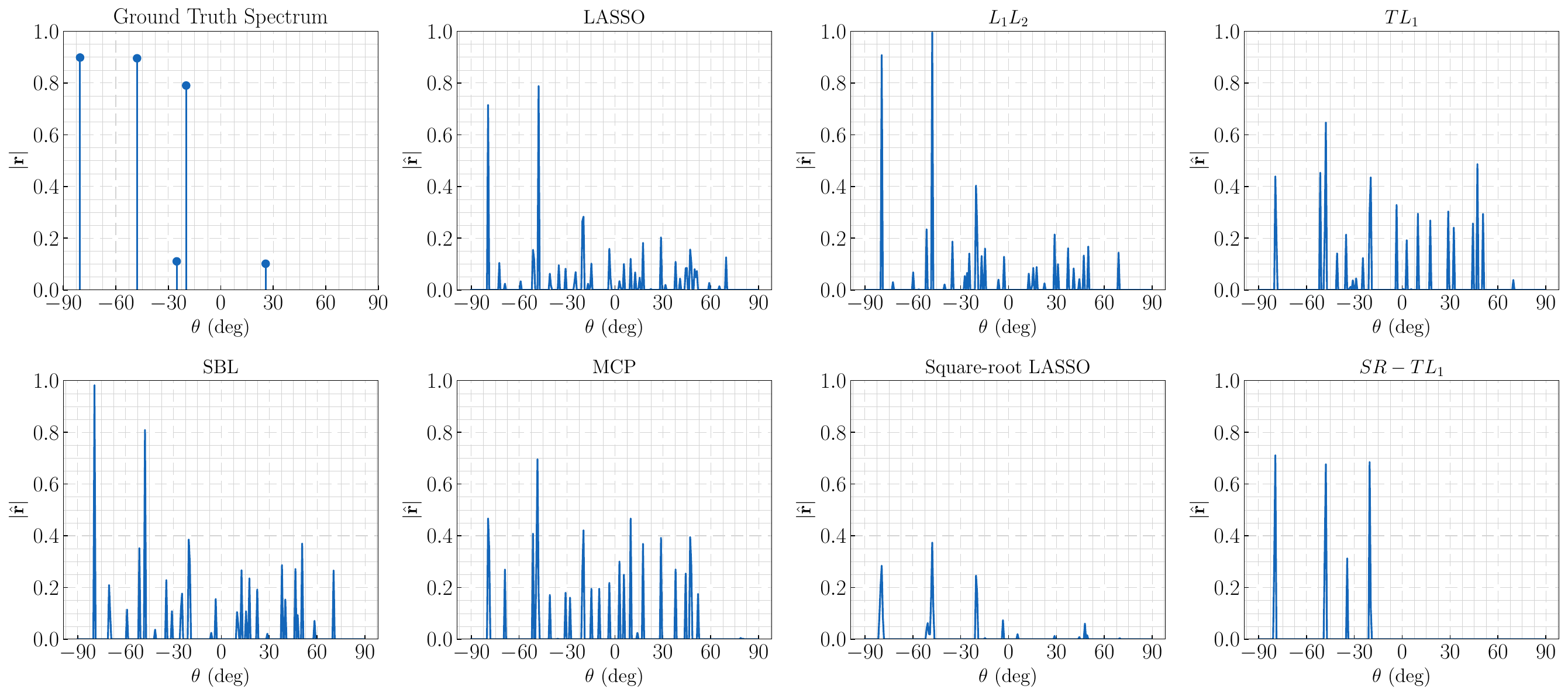}%
        }
        \caption{High noise variance setting with $\sigma^{2} = 1$ and SNR = 3.4 dB.}
        \label{sample_spectra_1_a}
    \end{subfigure}
    \vspace{0.1cm}
    
    \begin{subfigure}{\textwidth}
        \centering
        \makebox[\textwidth][c]{%
          \includegraphics[width=16cm, height=7cm, keepaspectratio=false]{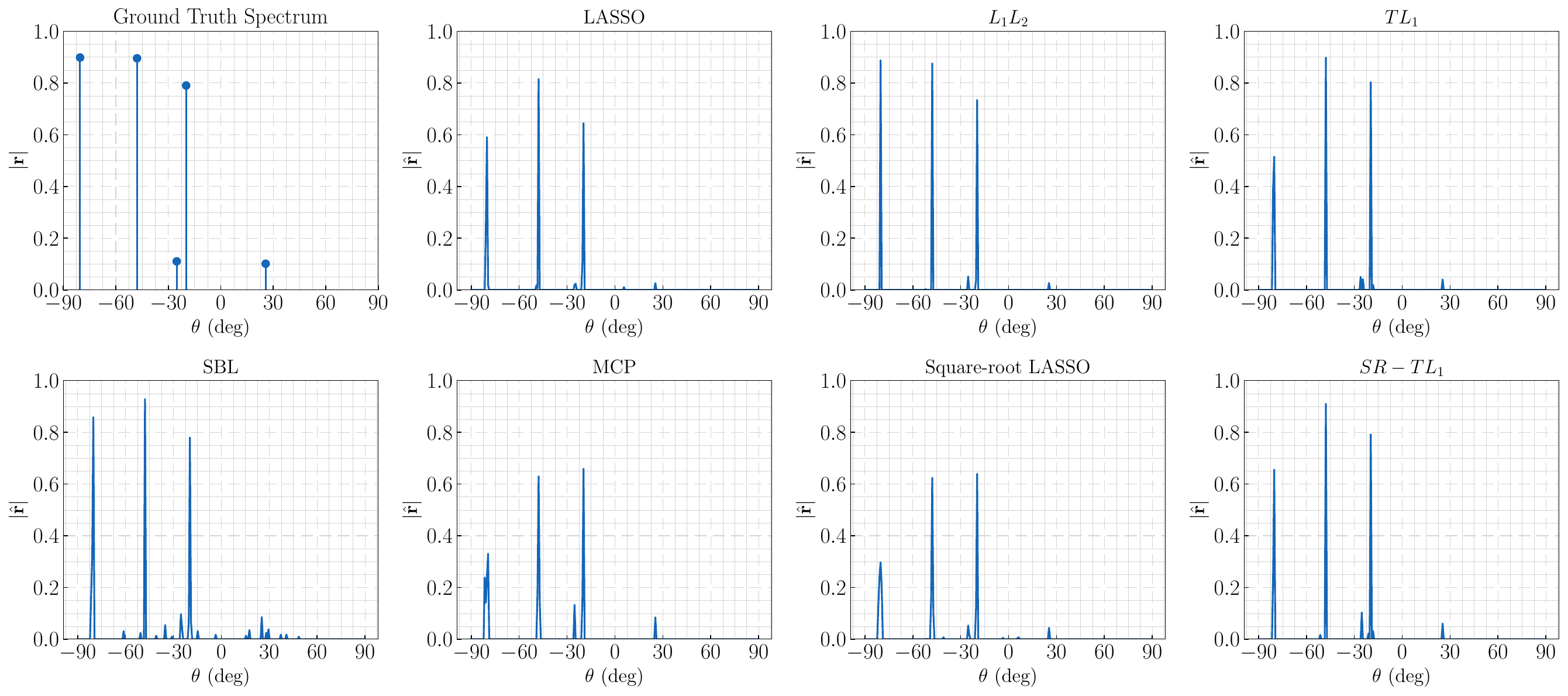}%
        }
        \caption{Medium noise variance setting with $\sigma^{2} = 10^{-2}$ and SNR = 23.4 dB.}
        \label{sample_spectra_1_b}
    \end{subfigure}
    \vspace{0.1cm}
    
    \begin{subfigure}{\textwidth}
        \centering
        \makebox[\textwidth][c]{%
          \includegraphics[width=16cm, height=7cm, keepaspectratio=false]{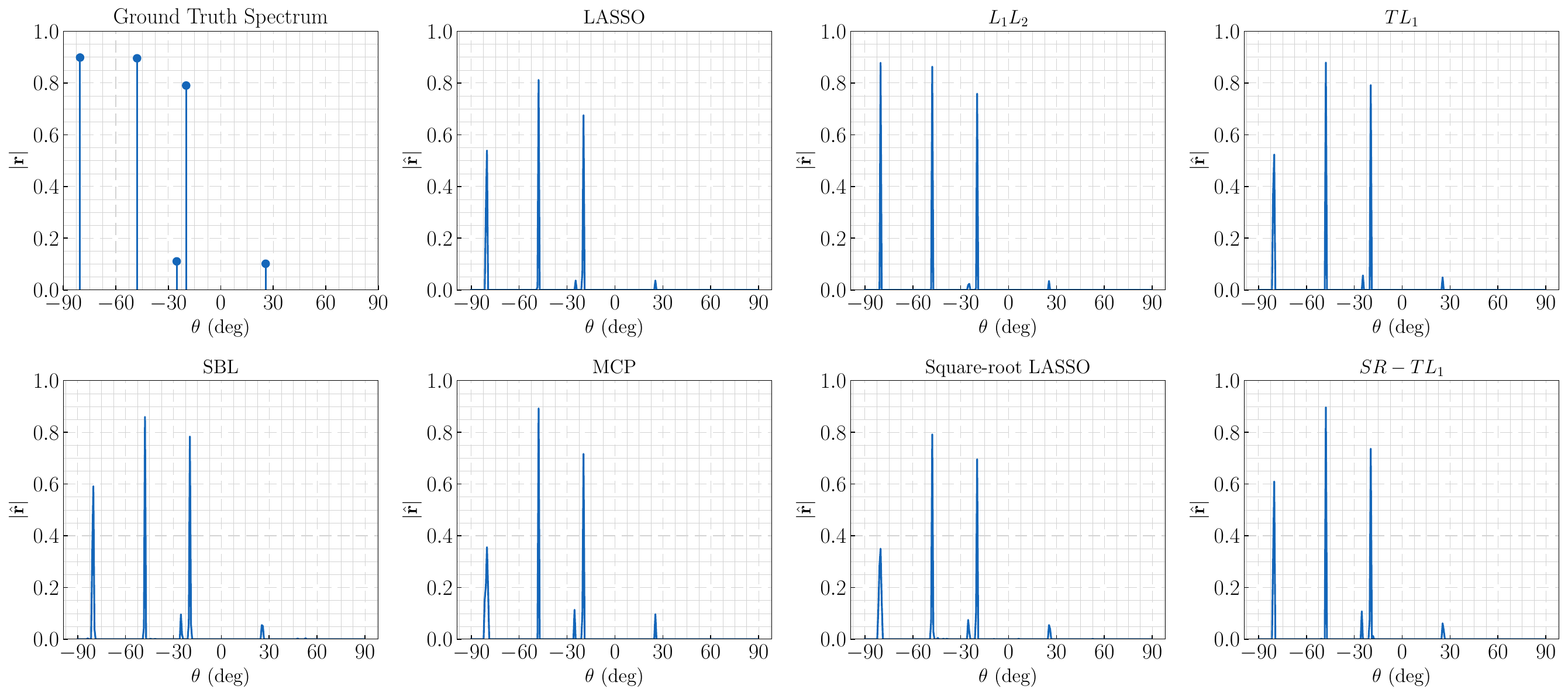}%
        }
        \caption{Low noise variance setting with $\sigma^{2} = 10^{-4}$ and SNR = 43.4 dB.}
        \label{sample_spectra_1_c}
    \end{subfigure}
    
    \caption{Sample 1 of reconstructed spectra for LASSO with $\kappa = 0.3$, $L_{1}L_{2}$ with $\kappa = 0.3$, $TL_{1}$ with $\kappa = 0.15$, MCP with $\kappa = 0.2$, square-root LASSO with $\kappa = 0.3$, and $SR\text{-}TL_{1}$ with $\kappa = 0.2$ for a noise variance level $\sigma^{2}\in \{1, 10^{-2}, 10^{-4}\}$.}
    \label{sample_spectra_1}
\end{figure}

\begin{figure}[!p]
    \centering
    \begin{subfigure}{\textwidth}
        \centering
        \makebox[\textwidth][c]{%
          \includegraphics[width=16cm, height=7cm, keepaspectratio=false]{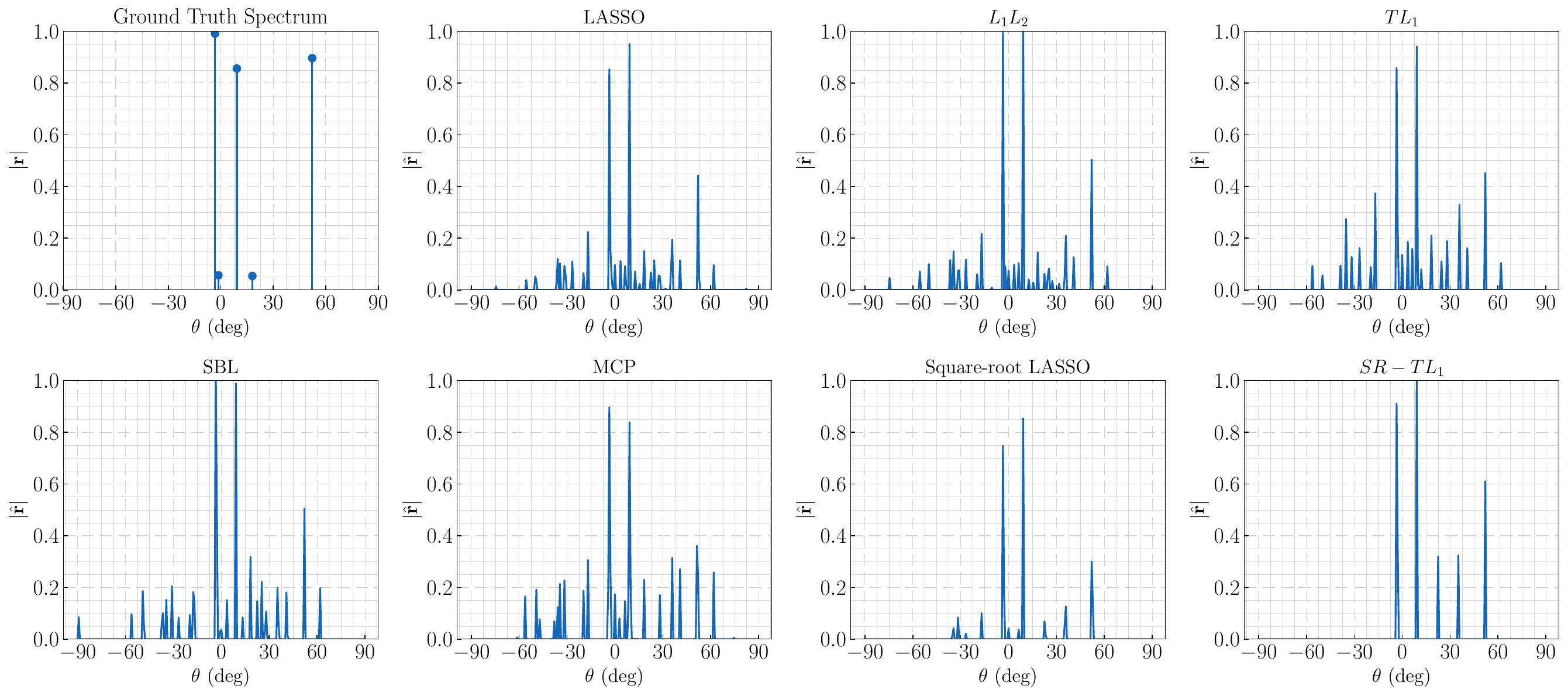}%
        }
        \caption{High noise variance setting with $\sigma^{2} = 1$ and SNR = 4.7 dB.}
        \label{sample_spectra_2_a}
    \end{subfigure}
    \vspace{0.1cm}
    
    \begin{subfigure}{\textwidth}
        \centering
        \makebox[\textwidth][c]{%
          \includegraphics[width=16cm, height=7cm, keepaspectratio=false]{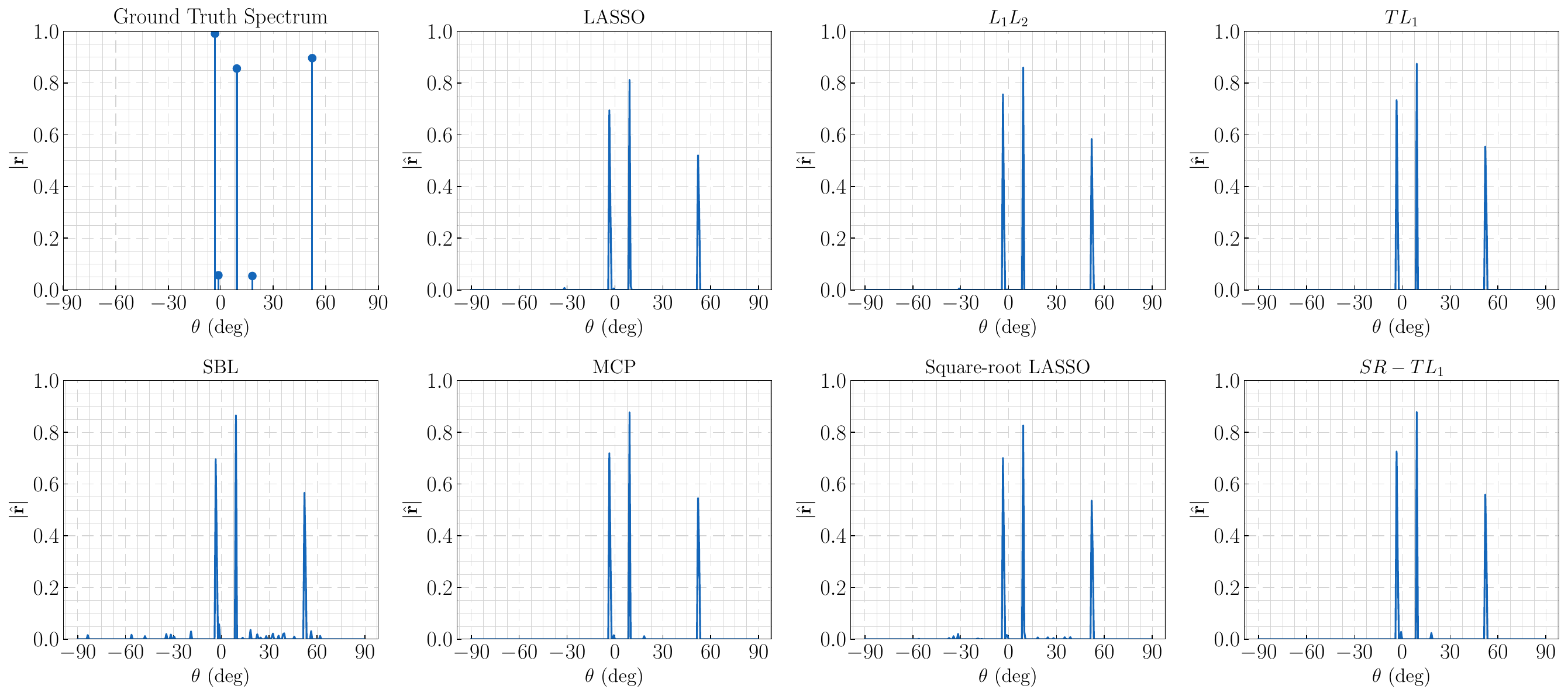}%
        }
        \caption{Medium noise variance setting with $\sigma^{2} = 10^{-2}$ and SNR = 24.7 dB.}
        \label{sample_spectra_2_b}
    \end{subfigure}
    \vspace{0.1cm}
    
    \begin{subfigure}{\textwidth}
        \centering
        \makebox[\textwidth][c]{%
          \includegraphics[width=16cm, height=7cm, keepaspectratio=false]{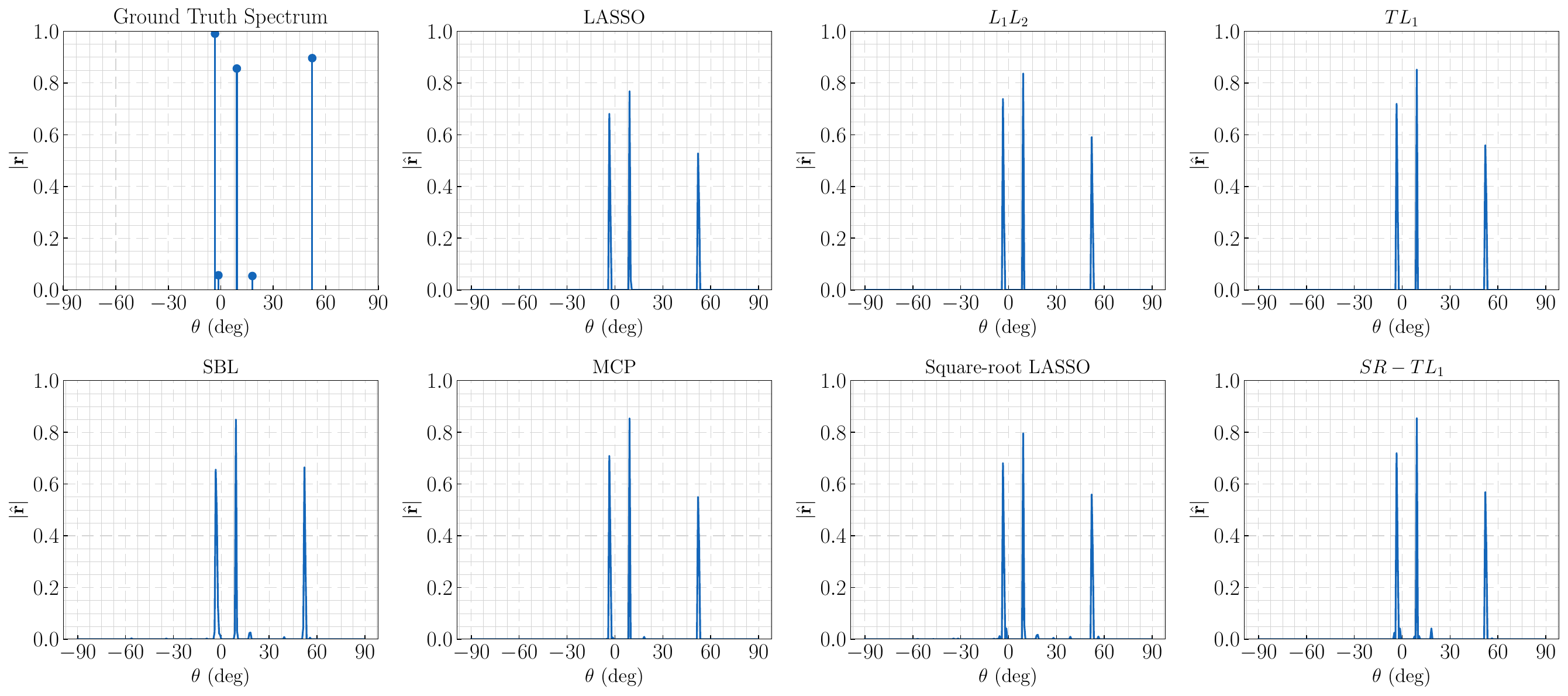}%
        }
        \caption{Low noise variance setting with $\sigma^{2} = 10^{-4}$ and SNR = 44.7 dB.}
        \label{sample_spectra_2_c}
    \end{subfigure}
    
    \caption{Sample 2 of reconstructed spectra for LASSO with $\kappa = 0.3$, $L_{1}L_{2}$ with $\kappa = 0.3$, $TL_{1}$ with $\kappa = 0.15$, MCP with $\kappa = 0.2$, square-root LASSO with $\kappa = 0.3$, and $SR\text{-}TL_{1}$ with $\kappa = 0.2$ for a noise variance level $\sigma^{2}\in \{1, 10^{-2}, 10^{-4}\}$.}
    \label{sample_spectra_2}
\end{figure}

As we can note in Fig. \ref{sr_lasso_detection_rate} and Fig. \ref{sr_tl1_detection_rate} the square-root LASSO and $SR\text{-}TL1_{1}$ both exhibit a smaller effective tuning range of the regularizer $\kappa$ compared to LASSO and $TL_{1}$, respectively. For example for $SR\text{-}TL_{1}$ for $\kappa\ge 0.3$ the detection rate drops significantly to under $P_{d}=0.1$ whereas for $TL_{1}$ at $\kappa = 0.7$ the detection rate is still above $P_{d}=0.5$. We can also note from Fig. \ref{sr_lasso_pfa} and Fig. \ref{sr_tl1_pfa} that the square-root frameworks exhibit a range of $\kappa$ values for which the false alarm rate remains approximately constant below a certain level over a wide noise variance interval. For example, for $\kappa = 0.2$ the $SR\text{-}TL_{1}$ framework is able to maintain a $P_{fa}$ under $0.04$ from $\sigma^{2}=10^{-4}$ up to $\sigma^{2}=1$. We can also see that although there are some values of $\kappa$ for $TL_{1}$ at which the $P_{fa}$ is small over a wide range, the same $\kappa$ values tend to correspond to low-value $P_{d}$ curves. Thus, for example, under $TL_{1}$ with $\kappa = 0.4$ we can maintain a $P_{fa}$ below $0.04$ from $\sigma^{2}= 10^{-4}$ up to around $\sigma^{2} = 10^{-0.5}$. If we now focus on the $P_{d}$ curve corresponding to $\kappa = 0.4$ for $TL_{1}$, we can see that the detection rate in the mid-to-high SNR region, which corresponds to approximately $\sigma^{2}=10^{-4}$ up to $\sigma^{2}=10^{-2}$, is approximately $0.7$, whereas $SR\text{-}TL_{1}$ has a $P_{d}$ around 0.9 over the same noise variance region. In other words, with $TL_{1}$, we can reduce the $P_{fa}$ in the high noise variance region by aggressively increasing the penalty regularizer $\kappa$. While this allows an increased suppression of noise in the low SNR regime, it also inevitably leads to the suppression of real signal components in the high SNR regime, thereby causing a decrease in the detection rate. We can also note that $SR\text{-}TL_{1}$ has a higher detection rate performance compared to square-root LASSO for a given $P_{fa}$ level. Thus for example, for $\kappa = 0.3$ square-root LASSO achieves a $P_{fa}$ performance under $0.04$ in the mid-to-high SNR region while achieving a detection performance of $P_{d} = 0.8$ in that same region compared again to $P_{d}=0.9$ for $SR\text{-}TL_{1}$. In terms of the RMSE performance, we can note from Fig. \ref{sbl_rmse} that SBL exhibits the best overall performance in the high SNR region for $\sigma^{2}\approx 10^{-4}$ with an RMSE value of around 0.3 deg. Although we should also note from Fig. \ref{sbl_pfa} that SBL has one of the highest $P_{fa}$ values of around $0.06$ in the same noise region compared to other baselines. Next to SBL, we can see from Fig. \ref{sr_lasso_rmse} and Fig. \ref{sr_tl1_rmse} that square-root LASSO, $TL_1$, and $SR\text{-}TL_{1}$ have roughly the same performance profile for $\sigma^{2}\le 10^{-2}$, whereas for $\sigma^{2}>10^{-2}$ square-root LASSO and $SR\text{-}TL_{1}$ begin to show a better performance compared to the other baseline methods for a given $P_{d}$ value, which is a consequence of their cleaner spectra in this low-SNR region compared to the other methods as indicated in the spectra samples in Fig. \ref{sample_spectra_1} and Fig. \ref{sample_spectra_2}.

\section{Conclusion}
\label{conclusion}
This work presented $SR\text{-}TL_{1}$ a non-convex sparse recovery framework for SMV DoA estimation under angular-dependent array imperfections that exhibits the noise-robust recovery performance of the square-root LASSO while also having a high recovery rate under highly coherent settings comparable to that of the $TL_{1}$ framework. We used the DCA and ADMM algorithms to provide simple closed-form update rules for solving the non-convex optimization problem. We provided a convergence guarantee of the resulting sequence of DCA iterates as well as an efficient implementation that leverages the low-rank nature of the dictionary. Experimental verification of the proposed framework across various noise variance levels and under a highly-coherent dictionary confirms its robust recovery performance compared to other state-of-the-art baselines.

\clearpage
\thispagestyle{figurepage}

\clearpage
\thispagestyle{figurepage}

\clearpage
\backmatter

\begin{appendices}
\section{Proof of Theorem \ref{th:conv}} \label{proof_theorem}

The proof is straightforward and relies on establishing that the sequence of iterates $\{\hat{\mathbf{c}}^{(t)}\}$ and the function $h(\mathbf{x}) = ||\mathbf{y}-\mathbf{D}\mathbf{x}||_{2} + \kappa R(\mathbf{x} ;\alpha)+ \frac{\tilde{\rho}}{2}||\mathbf{x}||_{2}^{2}$, where $
R(\mathbf{x}; \alpha) = \sum_{n=1}^{N} \frac{(\alpha + 1)|x(n)|}{\alpha + |x(n)|}$, meet three conditions in the ABS theorem \cite{attouch2013convergence}, namely 

\begin{enumerate}
\renewcommand{\labelenumi}{\roman{enumi})}
\item \emph{Sufficient decrease condition:} There is a $\zeta_{1}>0$ such that 
\begin{align}
h(\hat{\mathbf{c}}^{(t+1)}) + \zeta_{1} ||\hat{\mathbf{c}}^{(t+1)}-\hat{\mathbf{c}}^{(t)}||_{2}^{2} \le h(\hat{\mathbf{c}}^{(t)}). 
\end{align} 
\item  \emph{Relative error condition:} There is a $\zeta_{2}>0$ and for each $t$ there is a $\mathbf{w}^{(t+1)} \in \partial h(\hat{\mathbf{c}}^{(t+1)})$ such that 
\begin{align}
||\mathbf{w}^{(t+1)}||_{2} \le \zeta_{2} ||\hat{\mathbf{c}}^{(t+1)}-\hat{\mathbf{c}}^{(t)}||_{2}.
\end{align}
\item \emph{Continuity condition:} There exists a subsequence $\{\hat{\mathbf{c}}^{(t_{j})}\}$ with $t_{j} \in \mathbb{N}$ such that for $j\rightarrow +\infty$
\begin{align}
\hat{\mathbf{c}}^{(t_{j})} \rightarrow \hat{\mathbf{c}} \quad \textrm{and} \quad h(\hat{\mathbf{c}}^{(t_{j})}) \rightarrow h(\hat{\mathbf{c}}). 
\end{align}
\end{enumerate}

We will first prove that the sequence of iterates $\{\hat{\mathbf{c}}^{(t)}\}$ is bounded. 
Evaluating the subproblem objective in \eqref{dca_subproblem} at $\hat{\mathbf{c}}^{(t)}$ and $\hat{\mathbf{c}}^{(t+1)}$  gives
\begin{align}
&f(\hat{\mathbf{c}}^{(t+1)}) - 2\operatorname{Re}\langle \nabla g(\hat{\mathbf{c}}^{(t)}), \hat{\mathbf{c}}^{(t+1)}\rangle
 \nonumber \\ 
&\le f(\hat{\mathbf{c}}^{(t)})- 2\operatorname{Re}\langle \nabla g(\hat{\mathbf{c}}^{(t)}), \hat{\mathbf{c}}^{(t)}\rangle.
\label{optim_ineq}
\end{align}
From the convexity of $g$, we additionally have
\begin{align}
g(\hat{\mathbf{c}}^{(t+1)}) \ge  g(\hat{\mathbf{c}}^{(t)}) + 2\operatorname{Re}\langle \nabla g(\hat{\mathbf{c}}^{(t)}), \hat{\mathbf{c}}^{(t+1)}-\hat{\mathbf{c}}^{(t)}\rangle \nonumber \\
\implies 2\operatorname{Re}\langle \nabla g(\hat{\mathbf{c}}^{(t)}), \hat{\mathbf{c}}^{(t+1)}-\hat{\mathbf{c}}^{(t)}\rangle \le g(\hat{\mathbf{c}}^{(t+1)})-g(\hat{\mathbf{c}}^{(t)}).
\label{g_convx}
\end{align}
Combining \eqref{optim_ineq} with \eqref{g_convx} we obtain
\begin{align}
f(\hat{\mathbf{c}}^{(t+1)})-f(\hat{\mathbf{c}}^{(t)}) &\le g(\hat{\mathbf{c}}^{(t+1)})-g(\hat{\mathbf{c}}^{(t)}) \nonumber \\
\implies h(\hat{\mathbf{c}}^{(t+1)}) &\le h(\hat{\mathbf{c}}^{(t)}).
\end{align}
Thus the sequence $\{h(\hat{\mathbf{c}}^{(t)})\}$ is bounded with bound $h(\hat{\mathbf{c}}^{(0)})$, i.e. $h(\hat{\mathbf{c}}^{(t)}) \le h(\hat{\mathbf{c}}^{(0)})$ for any $t\in\mathbb{N}$ . We additionally have for any $\mathbf{x} \in \mathbb{C}^{N}$
\begin{align}
h(\mathbf{x}) \ge \frac{\tilde{\rho}}{2}||\mathbf{x}||_{2}^{2},
\end{align}
from which
\begin{align}
\lim_{||\mathbf{x}||_{2} \rightarrow +\infty} h(\mathbf{x}) \rightarrow +\infty,
\end{align}
and thus $h$ is coercive. We can therefore conclude from the coercivity of $h$ and the boundedness of $\{h(\hat{\mathbf{c}}^{(t)})\}$ that $\{\hat{\mathbf{c}}^{(t)}\}$ is bounded with bound $C$ given by
\begin{align}
C = \sqrt{\frac{2h(\hat{\mathbf{c}}^{(0)})}{\tilde{\rho}}}.
\end{align}
Since $h$ a continuous function, it follows that the \emph{continuity condition} in iii) is directly satisfied. For what follows, we define the closed ball $\mathcal{B}_{C} = \{\mathbf{x} \in \mathbb{C}^{N}: ||\mathbf{x}||_{2}\le C\}$.
\par Next we will establish that $h$ satisfies condition i). We have that for any $t \in \mathbb{N}$ 
\begin{align}
\hat{\mathbf{c}}^{(t)} \in \mathcal{B}_{C} \implies |\hat{\mathbf{c}}^{(t)}(n)| \le C \quad n=1, 2, \hdots N.
\end{align} 
Moreover, setting $\mathbf{x}(n) = \Re(\mathbf{x}(n)) + j\Im(\mathbf{x}(n))$, the function $g$ in the DC decomposition in \eqref{decompositon_r2} can be written as 
\begin{align}
g(\mathbf{x}) = \sum _{n=1}^{N} \tilde{g}(\mathbf{x}_{R}(n),\mathbf{x}_{I}(n) ),
\end{align}
where $\tilde{g}$ is a radial function given by 
\begin{align}
\tilde{g}(x,y) = \phi(p), \quad \phi(p) = \kappa'\frac{p^2}{\alpha+p}, \quad p=\sqrt{x^{2}+y^2}.
\end{align}
Defining $\mathbf{p} = [x,y]^{T}$, it can be easily verified that the Hessian of $\tilde{g}$ is given by
\begin{align}
\nabla ^{2}\tilde{g} = \phi''(p)\frac{\mathbf{p}\mathbf{p}^{T}}{p^{2}}+\frac{\phi'(p)}{p}\big ( \mathbf{I}- 
\frac{\mathbf{p}\mathbf{p}^{T}}{p^{2}} \big ),
\end{align}
where 
\begin{align}
\phi'(p) = \kappa'\frac{p(2\alpha+p)}{(\alpha+p)^2},\qquad
\phi''(p) = \frac{2\kappa'\alpha^2}{(\alpha+p)^3}.
\end{align}
The eigenvalues of $\nabla ^{2}\tilde{g}$ are additionally given by $\lambda_{1}(p) = \phi''(p)$ and $\lambda_{2}(p) = \phi'(p)/p$. On the set $p\in [0, C]$ we have $\lambda_{1}(p), \lambda_{2}(p) > 0$, from which $\tilde{g}$ is locally strongly convex. Since $g$ is the sum of locally strongly convex functions, there is a $\tilde{\zeta}_{1}>0$ such that for $\mathbf{x}_{1},\mathbf{x}_{2} \in \mathcal{B}_{C}$ the following holds
\begin{align}
g(\mathbf{x}_{2}) \ge g(\mathbf{x}_{1}) + 2 \Re\langle \nabla g(\mathbf{x}_{1}), \mathbf{x}_{2}-\mathbf{x}_{1}\rangle + \frac{\tilde{\zeta_{1}}}{2}||\mathbf{x}_{2}-\mathbf{x}_{1}||_2^2.
\end{align}
Setting $\mathbf{x}_{2} = \mathbf{c}^{(t+1)}$ and $\mathbf{x}_{1} = \mathbf{c}^{(t)}$ we obtain 
\begin{align}
2\Re\langle \nabla g(\hat{\mathbf{c}}^{(t)}), \hat{\mathbf{c}}^{(t+1)}-\hat{\mathbf{c}}^{(t)}\rangle
&\le g(\hat{\mathbf{c}}^{(t+1)}) - g(\hat{\mathbf{c}}^{(t)})  -\frac{\tilde{\zeta_{1}}}{2}||\hat{\mathbf{c}}^{(t+1)}-\hat{\mathbf{c}}^{(t)}||_2^2.
\label{g_stronglycvx}
\end{align}
Substituting \eqref{g_stronglycvx} into \eqref{optim_ineq} and rearranging we obtain
\begin{align}
h(\hat{\mathbf{c}}^{(t+1)}) + \frac{\tilde{\zeta}_{1}}{2} ||\hat{\mathbf{c}}^{(t+1)}-\hat{\mathbf{c}}^{(t)}||_{2}^{2} \le h(\hat{\mathbf{c}}^{(t)}).
\end{align} 
Thus $h$ satisfies the \emph{sufficient decrease condition}.
\par To prove ii), we can first note that since $g$ is continuously differentiable, the subdifferential of $h = f - g$ is given by $\partial h(\mathbf{x}) = \partial f(\mathbf{x}) - 2\nabla g(\mathbf{x})$. Additionally, the optimality condition for the convex problem in \eqref{dca_subproblem} is given by
\begin{align}
\mathbf{0} \in \partial f(\hat{\mathbf{c}}^{(t+1)})-2\nabla g(\hat{\mathbf{c}}^{(t)}) \implies 2\nabla g(\hat{\mathbf{c}}^{(t)}) \in \partial f(\hat{\mathbf{c}}^{(t+1)}),
\label{optimality}
\end{align}
from which
\begin{align}
2\nabla g(\hat{\mathbf{c}}^{(t)})- 2\nabla g(\hat{\mathbf{c}}^{(t+1)}) \in \partial f(\hat{\mathbf{c}}^{(t+1)})- 2\nabla g(\hat{\mathbf{c}}^{(t+1)}).
\end{align}  
Defining $\mathbf{w}^{(t+1)} = 2\nabla g(\hat{\mathbf{c}}^{(t)})- 2\nabla g(\hat{\mathbf{c}}^{(t+1)})$, we thus have
\begin{align}
\mathbf{w}^{(t+1)} \in \partial h(\hat{\mathbf{c}}^{(t+1)}).
\end{align}
Additionally, $\nabla g$ is continuously differentiable over the bounded compact set $\mathcal{B}_{C}$, from which the function $\nabla g$ is Lipschitz continuous over $\mathcal{B}_{C}$. Thus there exists a constant $\tilde{\zeta}_{2}$, such that for $\mathbf{x}_{1}, \mathbf{x}_{2} \in \mathcal{B}_{C}$ we have
\begin{align}
||\nabla g(\mathbf{x}_{2}) - \nabla g(\mathbf{x}_{1})||_2 \le \tilde{\zeta}_{2} ||\mathbf{x}_{2}-\mathbf{x}_{1}||_2,
\end{align}
and in particular for $\mathbf{x}_{2} = \hat{\mathbf{c}}^{(t+1)}$ and $\mathbf{x}_{1}=\hat{\mathbf{c}}^{(t)}$ we have 
\begin{align}
||\mathbf{w}^{(t+1)}||_2 \le 2\tilde{\zeta}_{2}||\hat{\mathbf{c}}^{(t+1)}-\hat{\mathbf{c}}^{(t)}||_2.
\end{align}
Thus the $\emph{relative error condition}$ in ii) is satisfied as well. 

\par Since $h$ is composed of semi‑algebraic functions (norms, absolute values, rational powers), $h$ satisfies the Kurdyka–Łojasiewicz (KL) property on $\mathbb{C}^{N}$. Thus from the ABS theorem \cite{attouch2013convergence} it follows that the bounded sequence of iterates $\{\hat{\mathbf{c}}^{(t)}\}$ converges to a critical point $\hat{\mathbf{c}}$ of $h$. In particular, for $t\rightarrow \infty$ we have $\hat{\mathbf{c}}^{(t)} \rightarrow \hat{\mathbf{c}}$ and $\hat{\mathbf{c}}^{(t+1)} \rightarrow \hat{\mathbf{c}}$, from which the optimality condition in \eqref{optimality} reduces to 
\begin{align}
\mathbf{0} \in \partial f(\hat{\mathbf{c}})-2\nabla g(\hat{\mathbf{c}}) \implies \mathbf{0} \in \partial h(\hat{\mathbf{c}}).
\end{align}
\end{appendices}

\end{document}